\documentclass[10pt, letterpaper, twoside]{article}
\usepackage[utf8]{inputenc}
\usepackage[T1]{fontenc}
\usepackage[bookmarks=false]{hyperref}
\usepackage[pdftex]{graphicx}
\usepackage{xcolor}
\usepackage{indentfirst}
\usepackage{amsmath}
\usepackage{amsfonts}
\usepackage{amssymb}
\usepackage{array}
\usepackage{comment}
\usepackage{mathtools} %\ell
\usepackage{bm}
\usepackage{upgreek}
\usepackage[width = 18cm, height = 23cm]{geometry}
\usepackage{lineno}
\usepackage{setspace}
\usepackage{cases}
\usepackage{multirow}
\usepackage{pbox}

\newcommand{\footremember}[2]{%
	\footnote{#2}
	\newcounter{#1}
	\setcounter{#1}{\value{footnote}}%
}

\newcommand{\blat}[1]{\ensuremath{\mathbf{#1}}}
\newcommand{\bgre}[1]{\ensuremath{\bm{#1}}}

\newcommand{\ba}{\blat{a}}
\newcommand{\bb}{\blat{b}}
\newcommand{\cc}{\blat{c}}
\newcommand{\bd}{\blat{d}}
\newcommand{\e}{\blat{e}}
\newcommand{\eeff}{\ensuremath{\varepsilon_{\mathrm{eq}}}}
\newcommand{\f}{\blat{f}}
\newcommand{\ft}{\ensuremath{f_{\mathrm{t}}}}
\newcommand{\feff}{\ensuremath{f_{\mathrm{eq}}}}

\newcommand{\Gt}{\ensuremath{G_{\mathrm{t}}}}
\newcommand{\bi}{\blat{i}}

\newcommand{\bj}{\blat{j}}

\newcommand{\m}{\blat{m}}
\newcommand{\M}{\blat{M}}
\newcommand{\n}{\blat{n}}
\newcommand{\oo}{\blat{o}}
\newcommand{\p}{\blat{p}}
\newcommand{\q}{\blat{q}}
\newcommand{\rr}{\blat{r}}
\newcommand{\s}{\blat{s}}
\newcommand{\seff}{\ensuremath{s_{\mathrm{eq}}}}
\newcommand{\bt}{\blat{t}}
\newcommand{\uu}{\blat{u}}
\newcommand{\vv}{\blat{v}}
\newcommand{\w}{\blat{w}}
\newcommand{\x}{\blat{x}}
\newcommand{\X}{\blat{X}}
\newcommand{\y}{\blat{y}}

\newcommand{\ttheta}{\bgre{\uptheta}}
\newcommand{\bphi}{\bgre{\upvarphi}}
\newcommand{\bomega}{\bgre{\upomega}}
\newcommand{\bvarepsilon}{\bgre{\upvarepsilon}}
\newcommand{\bchi}{\bgre{\upchi}}
\newcommand{\bxi}{\bgre{\upxi}}
\newcommand{\bzeta}{\bgre{\upzeta}}
\newcommand{\bvarphi}{\bgre{\upvarphi}}
\newcommand{\bgamma}{\bgre{\upgamma}}

\newcommand{\bsigma}{\bgre{\upsigma}}

\newcommand{\bmu}{\bgre{\upmu}}

\newcommand{\VRVE}{\ensuremath{V_0}}

\newcommand{\dd}[1]{\mathrm{\,d}\hspace{0.05em}#1}
\newcommand{\llevicivita}{\ensuremath{\bm{\mathcal{E}}}}
\newcommand{\levicivita}{\ensuremath{\mathcal{E}}}
\newcommand{\RVE}{\ensuremath{\mathrm{RVE}}}

\usepackage[makeroom]{cancel}

\usepackage[style=numeric,sorting=nyt,backend=biber, uniquename=false ,maxbibnames=30, maxcitenames=2]{biblatex}
\usepackage{empheq}

\makeatletter
\g@addto@macro\bfseries{\boldmath}
\makeatother

\begin{document}
%\linespread{1.0}
\pagestyle{plain}
\pagenumbering{arabic}
%\doublespacing
%\linenumbers
	
\title{Homogenization of discrete mesoscale model of concrete for coupled mass transport and mechanics by asymptotic expansion}
\author{Jan Eliáš\footremember{Brno}{Brno University of Technology, Faculty of Civil Engineering, Brno, Czechia}\footremember{email}{Corresponding author: jan.elias@vut.cz} \and Gianluca Cusatis\footremember{Northwestern}{Northwestern University, Department of Civil and Environmental Engineering, Evanston, IL USA}}
\date{}

\maketitle 

\section*{Abstract}

Mass transport phenomenon in concrete structures is strongly coupled with their mechanical behavior. The first coupling fabric is the Biot's theory according to which fluid pressure interacts with solid stress state and volumetric deformation rate of the solid induces changes in fluid pressure. Another coupling mechanism emerges with cracks which serve as channels for the fluid to flow through them and provide volume for fluid storage. Especially the second coupling mechanism presents a~challenge for numerical modeling as it requires detailed knowledge about cracking process. Discrete mesoscale mechanical models coupled with mass transport offer simple and robust way to solve the problem. On the other hand, however, they are computationally demanding. 

In order to reduce this computational burden, the present paper applies the asymptotic expansion homogenization technique to the coupled problem to deliver (i) continuous and homogeneous 
description of the macroscopic problem which can be easily solved by the finite element method, (ii) discrete and heterogeneous mesoscale problem in the periodic setup attached to each integration point of the macroscale along with (iii) equations providing communication between these two scales. The transient terms appear at the macroscale only, as well as the Biot's coupling terms. The coupling through cracking is treated at the mesoscale by changing conductivity of the conduit elements according to the mechanical solution, otherwise the two mesoscale steady state problems are decoupled and can be therefore solved in a~sequence. This paper presents verification studies showing performance of the homogenized solution. Further improvement is achieved by pre-computing the initial linear mesoscale solution and adaptively replacing it by the full nonlinear one only at integration points that fulfill Ottosen's stress-based criterion indicating deviation from linearity.

\section{Introduction \label{sec:intro}}

Reliable, safe and efficient design of concrete structures demands robust numerical models capable to predict structural behavior under various external conditions, including post-critical stages. Coupled multi-physical phenomena are being commonly investigated in design and assessment of long term behavior and durability of concrete structures. One of the most frequent coupled analysis is a~problem of mechanics and mass transport, which is the subject of investigation in the present paper. 

Coupled mechanics and mass transport in porous quasibrittle medium is nowadays almost exclusively solved with a~help of complex numerical models. Coupling is provided by the Biot's theory~\parencite{Bio41} where fluid pressure depends on volumetric deformation rate and solid stress state is affected by the fluid pressure. In addition, there is the a~coupling source due to presence of cracks which create additional volume for fluid storage and increase material permeability. One can find several coupled simulation methods treating the material as homogeneous, such as X-FEM~\parencite{RenYou21,WanHua-20} or phase-field~\parencite{MieMau-15,WilLan-16} techniques. The homogeneous approaches suffer from oversimplified representation of cracks and consequently less accurate capturing of the coupling effects.  

Fracture process in quasibrittle materials is a~complex and challenging problem. Research has demonstrated that it is advantageous to use mesoscale techniques that, by capturing features of the internal structure, provide accurate insight into the cracking state~\parencite{SalMat-16,YilMil17}. Thereby, mesoscale models provide better estimation of flux through the cracks, i.e.,~more accurate description of the coupling effects. There is, however, a~price paid for the quality of the results in a~form of computational burden associated with mesoscale models. A~partial reduction of this obstacle can be achieved by using discrete models where each numerical particle corresponds to a~large inclusion with a~surrounding matrix~\parencite{CusPel-11,AlnPel-19,EliVor20}. This not only drastically reduces the number of free parameters of the displacement field (called Degrees of Freedom or DoF hereinafter), it also brings several other advantages such as natural description of oriented and discontinuous cracks and simple vectorial form of constitutive equations~\parencite{BolEli-21}. On the other hand, there is also a~disadvantage in a~limited range of achievable Poisson's ratio of the discrete system~\parencite{CusPel-11,Eli20,BolEli-21}.

The mechanical discrete models have been extended by coupled multi-physical phenomena in the last two decades, the focus is placed on mass transport in particular. Initially, the transport problem was solved on the same discrete structure as mechanics, which brought difficulties in coupling flow with crack openings as the cracks were perpendicular to the conduit elements~\parencite{BolBer04}.  The pioneering paper of \textcite{Gra09} brought concept of a~dual lattice with aligned cracks and conduit elements allowing to accurately reflect the crack opening effect on the flow. The dual lattice approach was later extended to 3D~\parencite{GraBol16}. Several other papers employ this concept, e.g., for simulation of cracking due to rebar corrosion~\parencite{FahWhe-17} or hydraulic fracturing~\parencite{UlvSun18,AthWhe-18,AsaPan-18}. \textcite{LiZho-18} apply the dual lattice concept to shale fracture and add effects of mechanical volumetric strain rate on the pressure. \textcite{SheLi-20} used the dual lattice concept to accurately simulate concrete thermal spalling.

Computational cost of coupled mesoscale discrete models is unfortunately still substantial and prohibits their application in design of structures and structural elements of a~real size. There are several known and widely used techniques available for reducing computational complexity of numerical models. The techniques applicable for mechanical discrete mesoscale models spans from (i) coarse graining~\parencite{Mul02,LalRez-18,AbdAln20}, (ii) adaptivity~\parencite{Eli16,RokPee-17,CorMat-20,CheCha-21}, (iii) model order reduction such as Proper orthogonal decomposition~\parencite{Sir87,CorDos-15,KerPas-12,CecZho-18} and (iv) homogenization. 

Homogenization of heterogeneous materials became a~classical technique allowing major reduction of computational cost~\parencite{BenLio-78}. It relies on two fundamental assumptions: (i) existence of a~representative volume element (RVE) or periodic unit cell, a~material volume containing complete information about material internal structure and its properties; (ii) scale separation, an~assumption that the size of such RVE is much smaller than the size of actual domain of the problem. 
For relatively simple cases, one can develop an~analytical solution of the RVE response. For most cases, non-trivial constitutive equations or complex material internal structures prevent a~closed-form solution. Therefore computational homogenization~\parencite{SmiBre-98,AlaGan-21,LebPon-21},  sometimes called FE$^2$, has become widely employed. RVE is attached to every integration point of the macroscopic continuous homogeneous model where it serves as a~``virtual experiment''. At every iteration, the RVE is appropriately loaded, a~boundary value problem at microscale is numerically solved, its results are collected and returned back to the macroscale. The present paper derives a~computational homogenization procedure for discrete coupled models using asymptotic expansion technique.    

A fundamental challenge in computational homogenization of quasibrittle materials is strain localization. \textcite{GitAsk-07} concluded that the assumption about the existence of RVE does not hold in presence of strain localization. Fortunately, several successful attempts were recently conducted towards solution of this problem. \textcite{KouGee-04} extended the homogenization scheme by incorporating the second order gradients of the macroscopic displacement field to regularize the RVE problem. The same research team~\parencite{CoeKou-12a,CoeKou-12b} proposed a~relaxed periodic boundary condition to allow formation of a~crack inside the RVE under arbitrary angle. Furthermore, they separated the microscale response into a~homogeneous ``bulk'' part and a~localization related part. Similar separation of the fine scale problem into localized and homogeneous parts as well as enrichment of the RVE boundary conditions were developed by \textcite{Ung13}. The homogeneous part accounts for both elastic deformation and diffused cracking while the localized part represents regions where strain localization takes place. A~mathematical basis for this extension relies on a~modification of Hill--Mandel principle. Further developments of these concepts can be found in Refs.~\parencite{OliCai-15,TurHoo-18,TanDon-19}. The strain localization problem is not addressed in this paper and is left open for further investigation.

The present paper extends work of \textcite{RezCus16} who developed homogenization approach for mechanical discrete models via asymptotic expansion. The same homogenization technique is applied also in Refs.~\parencite{RezAln-19,RezZho-17}. Analogous strategy is taken in Ref.~\parencite{EliYin-22} to homogenize the mass transport problem in a~discrete setup. The present paper brings a~unifying technique by applying the asymptotic expansion homogenization to the coupled problem of mass transport and momentum balance~\parencite{AurBou-09}; the previously derived decoupled derivations appear as special cases of the coupled one developed here. The present homogenization approach is capable to provide reasonable accurate predictions at a~reduced computational cost and much simpler pre-processing. Final parts of the paper demonstrate capabilities of the homogenized solution by comparing the \emph{full} and \emph{homogenized} models with a~help of multiple examples.   

\section{Discrete model of the coupled problem \label{model}}

The asymptotic expansion homogenization developed further in this study can be generally applied to any discrete model with fixed connectivity between particles. The most prominent representant of this type of models is arguably the Lattice Discrete Particle Model (LDPM)~\parencite{CusPel-11,CusMen-11}, which has been applied to various loading scenarios including static-like loading or highly dynamical events (e.g.,  projectile penetration~\parencite{SmiCus-14,FenSon-18}). The model was originally formulated for plain and reinforced concrete but soon adapted also for short fiber reinforcement~\parencite{JinBur-16,CheWei-21}. It has been extended to simulate multi-physical coupled phenomena such as Alkali-Silica reaction~\parencite{AlnCus-13,RezAln-19} or thermal spalling~\parencite{SheLi-20}. LDPM has been extensively validated using large sets of experiments. The wide acceptance of LDPM makes it an~ideal model for the homogenization. The present paper therefore develops the homogenization for LDPM, even though a~slightly modified and simplified model is used in the verification examples.

\begin{figure}[!b]
\centering
\includegraphics[width=\textwidth]{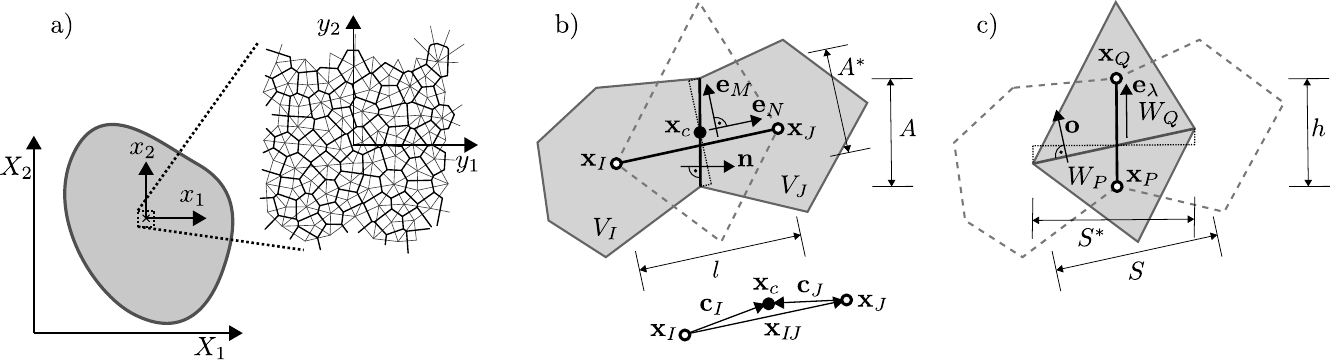}
\caption{2D sketch of the homogenization setup: a) different reference systems, a~continuous macroscale problem and a~discrete periodic microscale problem; b) a~mechanical element connecting two rigid particles; c) a~conduit element along the contact between two particles. \label{fig:2Dsketch}}
\end{figure}

The model geometry is based on a~tessellation of the computational domain into simplices (tetrahedrons in three dimensions). These tetrahedrons serve as control volumes, basic units for the mass transport problem in a~fully saturated medium as the mass balance within these control volumes is preserved. The pressure degree of freedom is assigned to each of the simplices, either to its centroid or to the center of its circumscribed sphere depending on the model type. The transport nodes are hereinafter denoted $P$ or $Q$ (Fig.~\ref{fig:2Dsketch}c). Each vertex of the simplex bears mechanical degrees of freedom (three translations and three rotations) and represents center of one mechanical rigid body/particle. For clarity, these mechanical nodes are denoted $I$ or $J$ (Fig.~\ref{fig:2Dsketch}b). The primary variables of the coupled problem are therefore the scalar field of pressure $p$ and vector fields of displacements $\uu$ and rotations $\ttheta$. The compressive fluid pressure is considered to be positive.
 
Each transport node $P$ at coordinates $\x_P$ with pressure degree of freedom $p^P$ is surrounded by faces (forming simplex called here the control volume) that connect it to other nodes. The connection between nodes $P$ and $Q$ has some area $S$, length $h=||\x_{PQ}||$  and contact direction $\e_{\lambda}  = \x_{PQ}/h$, where $\x_{PQ}=\x_Q-\x_P$. The situation is sketched in 2D in Fig.~\ref{fig:2Dsketch}c. Estimation of pressure gradient between two nodes yields
\begin{align}
g = \nabla p \cdot \e_{\lambda} \approx \frac{p^Q-p^P}{h} \label{eq:presgrad}
\end{align} 
The transport face normal $\oo$ (see Fig.~\ref{fig:2Dsketch}c) might not in general be  parallel to the contact vector $\e_{\lambda}$. Projected area $S^{\star}=S \oo\cdot \e_{\lambda}$ is introduced to account for the directional mismatch when the total flux is computed later in Eq.~\eqref{eq:balanceTransport}. 

The mechanical node $I$ is connected to the surrounding mechanical nodes $J$. The contact faces form a~polyhedron as shown in Fig.~\ref{fig:2Dsketch}b. 
The strain and curvature at the contact between bodies $I$ and $J$ read
\begin{align}
\varepsilon_{\alpha} &= \frac{1}{l}\left(\uu^J-\uu^I+\llevicivita:(\ttheta^J \otimes \cc_J - \ttheta^I \otimes \cc_I) \right)\cdot \e_{\alpha} &
\chi_{\alpha} &= \frac{1}{l}\left(\ttheta^J - \ttheta^I \right)\cdot \e_{\alpha} \label{eq:strain}
\end{align}
where $\cc$ is a~vector connecting particle governing node with the integration point $\x_c$ at the contact face (see Fig.~\ref{fig:2Dsketch}b), $\llevicivita$ is Levi-Civita permutation symbol and $\e_{\alpha}$ are local normal and two tangential directions ($\alpha\in\left\{N,\,M,\,L\right\}$), respectively. The length of the contact is $l=||\x_{I\! J}||$ and the contact direction $\e_{N}=\x_{I\! J}/l$ where  $\x_{I\! J} = \x_J-\x_I$. The contact direction $\e_{N}$ can in general be different from the true face normal $\n$. The contact area $A$ is therefore again projected as $A^{\star}=A \e_N\cdot\n$. Equations~\eqref{eq:presgrad} and \eqref{eq:strain} are the compatibility equations of the model providing strain-like variables from primary fields.  

The second set of equations are constitutive equations which provide stress-like variables. These are flux scalar, $j$, traction vector, $\bt$, and couple traction vector, $\m$
\begin{align}
j &= f_{j} (p_{\lambda},g,\delta_{\lambda}) = -\lambda(p_{\lambda}, \delta_{\lambda}) g & \bt &= f_s\left(\bvarepsilon\right) - bp_a\e_N   & \m = f_m\left( \bchi \right)  \label{eq:LDPMconst}
\end{align}
The first equation expresses linear dependence of the flux, $j$, on the pressure gradient, $g$, while the permeability coefficient, $\lambda$, is governed by an~average normal crack opening, $\delta_{\lambda}$, and an~average pressure, $p_{\lambda}$, in the element. 
Several possible formulations are being used to describe effects of cracks \parencite{AsaPan-18,AthWhe-18,LiZho-18} or pressure \parencite{Gen80} on material permeability in the literature. The second constitutive equation defines another coupling between transport and mechanics as the total traction, $\bt$, becomes dependent on the fluid pressure according to Biot's theory~\parencite{Bio41,DetChe95}, $b$ is a~material parameter called Biot coefficient and $\s = f_s\left( \bvarepsilon \right)$ is the vector of traction in the solid. The pressure $p_a$ is the weighted average pressure from control volumes surrounding the mechanical element. The third constitutive equation assumes that the couple traction, $\m$, is dependent solely on the curvature, $\bchi$, i.e.,~decoupled from the transport part of the model. The homogenization procedure is in principle independent of the choice of the functions $\lambda$, $f_s$ and $f_m$, but their dependency on other variables would require adjustments of Eqs.~\eqref{eq:stress1} .

Finally, balance equations are assembled. The balance of linear and angular momentum of particle $I$ read
\begin{align}
V\rho \ddot{\uu}^I + \M_{u\theta} \cdot \ddot{\ttheta}^I - V \bb &= \sum_J A^{\star}t_{\alpha}\e_{\alpha} 
&
\M_{\theta} \cdot \ddot{\ttheta}^I + \M_{u\theta}^T \cdot \ddot{\uu}^I &= \sum_J A^{\star}\left[\w + m_{\alpha}\e_{\alpha}\right] \label{eq:LDPMbalance}
\end{align}
where $\w = \llevicivita:(\cc_I\otimes \bt) =t_{\alpha}\llevicivita:(\cc_I\otimes \e_{\alpha})$ is the moment of traction with respect to the mechanical node $\x_I$. $\M_{\theta}$ and $\M_{u\theta}$ are the moment of inertia tensors defined as
\begin{align} \label{eq:inertia}
\M_{\theta} & = \rho \int_V  \rr \cdot \rr\, \bm{1}  - \rr \otimes \rr \dd{V} &
\M_{u\theta} = -\M_{u\theta}^T &= \rho V \llevicivita\cdot \rr_{0}
\end{align}
with $\rr$ being a~vector from the particle governing node $\x_I$ to any point within the particle and $\rr_0$ being a~vector from $\x_I$ to the particle centroid; $\bm{1}$ is the second order identity tensor. For special case $\rr_{0}=\bm{0}$ when the governing node of the particle coincides with its centroid, $\M_{u\theta}=\bm{0}$. Such situation corresponds to a~specific model geometry (for example the one obtained by centroidal Voronoi tessellation) but not to the general case of the described model. 

Note, that the constitutive equations are computed in a~rotated local reference system $\left[ \e_{N}, \e_M, \e_L\right]$ and therefore both traction components need to be projected back to the $\x$ reference system in the equilibrium equations. The volume of the particle is $V$, $\rho$ is the solid  density and $\bb$ represents the external load. If needed, a~moment source term can be easily added to the balance of angular momentum.

The mass balance equation for a~fully saturated medium is established for each control volume $P$ separately. It is assumed that the liquid is slightly compressible with bulk modulus $K_w$. The pressure dependent fluid density becomes~\parencite{LiZho-18}
\begin{align}
\rho_w = \rho_{w0}\left(1+\frac{p_{\lambda}-p_0}{K_w}\right)
\end{align}
where $p_0$ is a~reference pressure and $\rho_{w0}$ is the fluid density under the reference pressure. According to the Biot's theory, we can express change of the mass of the liquid in the reference volume $W$ as~\parencite{LiZho-18}
\begin{align}
\rho_{w0}\left(3 b \dot{\varepsilon}_{V,\mathrm{eff}}  + \frac{\dot{p}_{\lambda}}{M_b}\right)W
\end{align} 
with $M_b$ being the Biot modulus (reciprocal of capacity, $c$) and $\varepsilon_{V,\mathrm{eff}}$ is the effective volumetric strain. In a~discrete setup, a~local volumetric strain is estimated as one third of the relative difference in a~tetrahedron volume due to displacements $\varepsilon_V = \left[W(\x+\uu)-W(\x)\right]/3W(\x)$. The effective volumetric strain is then obtained by removing the volume of cracks from the volumetric strain 
\begin{align} \label{eq:volum_strain_effective}
\varepsilon_{V,\mathrm{eff}} = \frac{W(\x+\uu)-W(\x) - W\sum_{Q\in W} v_c }{3W(\x)} = \varepsilon_V - \frac{1}{3}\sum_{Q\in W} v_c
\end{align}
with $v_c$ being the relative crack volume within a~conduit element connecting nodes $P$ and $Q$ evaluated as the total crack volume in the element, $V_c$, divided by element volume $W_P+W_Q$. The total crack volume $V_c$ reads
\begin{align}
V_c = \sum_{\mathrm{neighb.}\ e} A_{f} \delta_{N}
\end{align} 
$A_{f}$ corresponds to fractions of areas of mechanical elements, $e$, connected to the conduit element and $\delta_{N}$ are their normal openings. Finally, change of the fluid mass in time due to cracks filled with fluid becomes
\begin{align}
W\frac{\partial \rho_w v_c}{\partial t} = W\left(\rho_{w} \dot{v}_c + v_c \rho_{w0}\frac{\dot{p}_{\lambda}}{K_w}\right)
\end{align}

Summing all the contributions together and adding also fluxes through the individual conduit elements and possible source term $q$ (positive when material flows into the volume), the discrete mass balance equation for control volume $P$ reads
\begin{align}
\sum_{Q\in W} \left[ S^{\star} j  - \rho_{w0} W \dot{v}_c \left(1 -b+\frac{p_{\lambda}-p_0}{K_{w}}\right)  -  \rho_{w0}W v_c\frac{\dot{p}_{\lambda}}{K_w}  \right] - \rho_{w0}\left(3 b \dot{\varepsilon}_V  + \frac{\dot{p}_{\lambda}}{M_b}\right) W - Wq= 0 \label{eq:balanceTransport}
\end{align}

\section{Separation of scales}

Two spatial variables are considered now for every point in the domain: the macroscopic, slow variable $\x$ and the microscopic, fast variable $\y$ (see Fig.~\ref{fig:2Dsketch}a). The following scale separation relationship holds
\begin{align}
\x &= \eta \y  \label{eq:scalesep}
\end{align} 
with $\eta$ being the separation of scales constant with properties $0<\eta\ll1$. The model appears continuous from the viewpoint of reference system $\x$ but discrete in the reference system $\y$ for sufficiently small $\eta$.

Moreover, another global reference system $\X$ that uniquely defines position in the continuous macroscopic space is introduced. System $\X$ has the same units as $\x$ but there is only one such system while infinitely many $\x$ reference systems are defined in all macroscopic spatial points. 

All the variables involving length are consider to be in the $\x$ reference system. One needs to transfer them into the $\y$  reference system according to the following transformation rules (the powers of $\eta$ reflect the power of the distance unit involved, $\tilde{\bullet}$ denotes variable $\bullet$ in $\y$ reference system) 
\begin{align}
\begin{aligned} \label{eq:scaling}
\x_{I\! J} &= \eta{\y}_{I\! J} & l &= \eta \tilde{l} & A^{\star} &= \eta^2 \tilde{A}^{\star} &
V &= \eta^3 \tilde{V} & \M_{u\theta} &= \eta^4 \tilde{\M}_{u\theta} &
\M_{\theta} &= \eta^5 \tilde{\M}_{\theta} \\ \x_{PQ} &= \eta{\y}_{PQ} & h &= \eta \tilde{h} & S^{\star} &= \eta^2 \tilde{S}^{\star} &
W &= \eta^3 \tilde{W} & \cc_J &= \eta\tilde{\cc}_J & 
\end{aligned}
\end{align}

The primary variables $p$, $\uu$ and $\ttheta$ are now considered to be approximated by several components, especially by the macroscopic component, $\bullet^{(0)}$, and the microscopic  component, $\bullet^{(1)}$. Expansion of the pressure and displacements is straightforward. Expansion of rotations is done according to \textcite{RezCus16} by assuming existence of some continuous displacement-like field $\bd$ the curl of which provides the rotations: $2\ttheta = \nabla \times \bd$. The expansion of the rotations is taken as curls of the expanded field $\bd$ with nabla operator transferred in accordance with Eq.~\eqref{eq:scalesep} into $\nabla = \nabla_X + \eta^{-1}\nabla_y$. The three asymptotic expansion equations read 
\begin{subequations} \label{eq:expansion}
\begin{align} 
p(\X,\y) &= \eta^{-1}p^{(-1)}(\X,\y) + p^{(0)}(\X,\y) + \eta p^{(1)}(\X,\y) + \dots \\
\uu(\X,\y) &= \uu^{(0)}(\X,\y) + \eta \uu^{(1)}(\X,\y) + \dots  \label{eq:expansion_u}\\
\ttheta(\X,\y) &= \eta^{-1}\bomega^{(-1)}(\X,\y) + \bomega^{(0)}(\X,\y) + \bphi^{(0)}(\X,\y)  + \eta \bphi^{(1)}(\X,\y) + \dots
\end{align}
\end{subequations} 
where $2\bomega^{(-1)} = \nabla_y \times \bd^{(0)}$,  $2\bomega^{(0)} = \nabla_y \times \bd^{(1)}$,  $2\bphi^{(0)} = \nabla_X \times \bd^{(0)}$ and  $2\bphi^{(1)} = \nabla_X \times \bd^{(1)}$. Note that the expansion of pressure should correspond to the expansion of the traction in mechanical models (Eq.~\ref{eq:traction_expansion}). Therefore, term $\eta^{-1}p^{(-1)}$ is included as well. All of the fields are assumed to be $y$-periodic over some Representative Volume Element (RVE) with periodic geometrical structure of the discrete model. It is also assumed that all these fast fields (or fluctuations) yield zero average over the RVE. 

From the viewpoint of the macroscopic spatial coordinate $\X$, neighboring mechanical nodes $I$ and $J$ are close to each other. According to \textcite{FisChe-07}, one can use the macroscopic gradient $\nabla_X$ at node $I$ to approximate the mechanical field variables at node $J$. The same Taylor expansion of pressure around node $P$ provides pressure estimation at neighboring node $Q$.
\begin{subequations} \label{eq:Taylorseries}
\begin{align}
p(\X_Q, \y_Q) &= p(\X_P, \y_Q) + \frac{\partial p(\X_P, \y_Q)}{\partial X_i} x_{i}^{PQ} + \frac{1}{2} \frac{\partial^2 p(\X_P, \y_Q)}{\partial X_i X_j} x^{PQ}_i x^{PQ}_j + \dots \\
\uu(\X_J,\y_J) &= \uu(\X_I,\y_J) + \frac{\partial \uu(\X_I,\y_J)}{\partial X_j} x^{I\! J}_j +  \frac{1}{2}\frac{\partial^2 \uu(\X_I,\y_J)}{\partial X_j X_k} x^{I\! J}_j x^{I\! J}_k + \dots \label{eq:Taylorseries_u}\\
\ttheta(\X_J, \y_J) &= \ttheta(\X_I,\y_J) + \frac{\partial \ttheta(\X_I,\y_J)}{\partial X_j} x^{I\! J}_j +  \frac{1}{2}\frac{\partial^2 \ttheta(\X_I,\y_J)}{\partial X_j X_k} x^{I\! J}_j x^{I\! J}_k + \dots
\end{align} 
\end{subequations}

The compatibility equations~\eqref{eq:presgrad} and \eqref{eq:strain} are now rewritten 
\begin{subequations} \label{eq:compatibileq}
\begin{align}
g &= \frac{1}{h}\left[p(\X_Q,\y_Q) - p(\X_P,\y_P)\right]\\
\varepsilon_{\alpha} &= \frac{1}{l}\left[\uu(\X_J,\y_J)-\uu(\X_I,\y_I)+\llevicivita:\left(\ttheta(\X_J,\y_J) \otimes \cc_J - \ttheta(\X_I,\y_I) \otimes \cc_I \right)\right]\cdot \e_{\alpha} \\
\chi_{\alpha} &= \frac{1}{l}\left[\ttheta(\X_J,\y_J) - \ttheta(\X_I,\y_I) \right]\cdot \e_{\alpha}
\end{align}
\end{subequations}
and the Taylor expansion~\eqref{eq:Taylorseries} as well as the asymptotic expansion~\eqref{eq:expansion} is substituted afterwards. Note that thanks to the Taylor expansion,  there is always the same $\X$ coordinate (either $\X_P$ or $\X_I$) denoting the RVE location. The equations must hold for any $\X$ coordinate (or any RVE), therefore we can drop it for sake of simplicity. As long as only a~single RVE is involved (constant $\X$ coordinate) the notation can be simplified $\bullet^{\alpha P} = \bullet^{(\alpha)}(\X,\y_P)$. For example, $\uu^{0I}$ means $\uu^{(0)}(\X,\y_I)$ hereinafter. 

The set of compatibility equations~\eqref{eq:compatibileq} with substituted \eqref{eq:expansion} and \eqref{eq:Taylorseries}  and scaled length variables \eqref{eq:scaling} reads
\begin{subequations} \label{eq:straindecomp}
\begin{align}
g & = \eta^{-2} g^{(-2)} + \eta^{-1} g^{(-1)} + g^{(0)} + \eta g^{(1)} + \dots\\
\varepsilon_{\alpha} &= \eta^{-1}\varepsilon^{(-1)}_{\alpha} + \varepsilon^{(0)}_{\alpha} + \eta\varepsilon^{(1)}_{\alpha} + \dots\\
\chi_{\alpha} &= \eta^{-2}\chi^{(-2)}_{\alpha} + \eta^{-1}\chi^{(-1)}_{\alpha} + \chi^{(0)}_{\alpha} + \dots
\end{align}
\end{subequations}
The individual components can be found in Eqs.~\eqref{eq:gechi_expansion} in Appendix~\ref{ap:strain_expansion}.

Next, the volumetric strain, $\varepsilon_V$, crack opening, $\delta_N$, and cracks volume density, $v_c$, are expanded. Volumetric strain, defined as one third of the relative change of the volume of the tetrahedron, is directly dependent on displacements $\uu$ according to Eq.~\eqref{eq:volum_strain_effective}. Since it is a~strain variable, its expansion must be the same as expansion of the strain. Normal crack opening $\delta_N$ is dependent on strain $\varepsilon_N$ by subtracting its elastic part and multiplying the result by length $l$. The relative crack volume $v_c$  is dependent on crack opening $\delta_N$ by multiplying it by the crack area $A_f$ and divided by the element volume. The expansions therefore read
\begin{subequations} \label{eq:expansion_volstr_cracks}	
\begin{align}
\varepsilon_V  &= \eta^{-1}\varepsilon^{(-1)}_V + \varepsilon^{(0)}_V  + \eta\varepsilon^{(1)}_V + \dots \label{eq:vol_strain_exp}\\
\delta_N  &= \delta^{(0)}_N + \eta\delta^{(1)}_N + \eta^2\delta^{(2)}_N + \dots \\
v_c  &= \eta^{-1}v^{(-1)}_c + v^{(0)}_c + \eta^1v^{(1)}_c + \dots
\end{align}
\end{subequations}

The stress-like variables (flux, traction, couple traction and moment of traction) are also expanded into 
\begin{subequations} \label{eq:stressdecomp}
\begin{align}
j & = \eta^{-2} j^{(-2)} + \eta^{-1} j^{(-1)} + j^{(0)} + \eta j^{(1)}+ \dots 
\\
\bt & = \eta^{-1} \bt^{(-1)} + \bt^{(0)} + \eta \bt^{(1)} + \dots \label{eq:traction_expansion}
\\
\m & = \m^{(0)} + \eta\m^{(1)} + \eta^2\m^{(2)} + \dots 
\\
\w & = \w^{(0)} + \eta\w^{(1)} + \eta^2 \w^{(2)} + \dots
\end{align}
\end{subequations}
It will be shown in Secs.~\ref{sec:zerop-1} and \ref{sec:rbm} that terms $j^{(-2)}$, $j^{(-1)}$, $\bt^{(-1)}$, $\m^{(0)}$, $\w^{(0)}$ as well as $g^{(-2)}$, $g^{(-1)}$, $\bvarepsilon^{(-1)}$, $\bchi^{(-2)}$, $\delta^{(0)}_{\lambda}$ and $p^{(-1)}$ are always zero. Therefore the first nonzero terms in expansion~\eqref{eq:stressdecomp} are expressed by components of the constitutive relations~\eqref{eq:LDPMconst} with the lowest $\eta$ powers
\begin{subequations}\label{eq:stressSecondTerms} 
	\begin{align}
	j^{(0)} = &-\lambda\left(p^{(0)}_{\lambda}, \eta\delta^{(1)}_{\lambda} \right) g^{(0)} \label{eq:stressSecondTermsFlux} \\
	\bt^{(0)} =  &f_s\left(\bvarepsilon^{(0)}\right) - p^{(0)}_a b\e_N \label{eq:stressSecondTermsTraction} \\
	\eta\m^{(1)} =  &f_m\left(\eta^{-1}\bchi^{(-1)}\right)\\
	\eta\w^{(1)} =  & t^{(0)}_{\alpha}\llevicivita:(\cc_I\otimes \e_{\alpha}) \label{eq:stressSecondTermsMomentOfTraction}
	\end{align}
\end{subequations}
These equations represent the actual constitutive equations of the discrete model. The second nonzero components of stress-like variables are developed via Taylor series in Eqs.~\eqref{eq:stress1} in Appendix~\ref{ap:stress_taylor}.

The source terms $q$ and $\bb$ (as well as other variables, e.q. Biot coefficient $b$, capacity $c$ or density $\rho$) might also be dependent on the primary or other fields. For example recent study~\parencite{RahCha-19} suggests that damage dependent Biot coefficient should be used for hydraulic fracturing simulations. In such cases, their expansion must be developed as well and added to the balance equations in the next section. It is assumed here, for sake of simplicity, that these variables are independent on the primary fields and have only one component of order $\approx\mathcal{O}(\eta^0)$. Reader interested in the described extension is referred to paper~\parencite{EliYin-22} where such expansion is developed for $q$ and $c$. 

\section{Balance equations}
The balance equations of individual particles or control volumes are now assembled in the $\x$ reference system using variables transformed into the $\y$ reference system. Starting with Eqs.~\eqref{eq:LDPMbalance} and \eqref{eq:balanceTransport}, assuming all material parameters are of order $\approx\mathcal{O}(\eta^0)$, transforming all the length variables from $\x$ to $\y$ reference system, and dividing everything by $\eta^3$, $\eta^3$ or $\eta^4$, respectively, the balance equations read
\begin{subequations} \label{eq:scaledBalanceAll}
\begin{align}
\rho_{w0}\tilde{W}\left(3 b \dot{\varepsilon}_V  + \frac{\dot{p}_{\lambda}}{M_b} \right) + \tilde{W}q &= \frac{1}{\eta}\sum_{Q\in W} \left[ \tilde{S}^{\star} j  - \eta \rho_{w0} \tilde{W}\dot{v}_c \left(1 - b + \frac{p_{\lambda}-p_0}{K_{w}}\right)- \eta\rho_{w0}\tilde{W} v_c\frac{\dot{p}_{\lambda}}{K_w}  \right] \label{eq:scaledBalanceFlow} 
\\
\tilde{V}\rho \ddot{\uu}^I +  \eta  \tilde{\M}_{u\theta} \cdot \ddot{\ttheta}^I -\tilde{V} \bb &= \frac{1}{\eta}\sum_J \tilde{A}^{\star} t_{\alpha} \label{eq:scaledBalanceForces} 
\\
\eta \tilde{\M}_{\theta}\cdot \ddot{\ttheta}^I + \tilde{\M}_{u\theta}^T \cdot \ddot{\uu}^I &= \frac{1}{\eta^2}\sum_J \tilde{A}^{\star}(\w + m_{\alpha} \e_{\alpha}) \label{eq:scaledBalanceMoments} 
\end{align}
\end{subequations}
The expansions developed in Eqs.~\eqref{eq:expansion}, \eqref{eq:straindecomp}, \eqref{eq:expansion_volstr_cracks} and \eqref{eq:stressdecomp} can now be substituted into the balance equations, which can be then decomposed into separate equation sets collecting the terms with corresponding powers of $\eta$.

\subsection{$\eta^{-3}$: zero pressure $\eta^{-1}p^{(-1)}$ \label{sec:zerop-1}}
The expansion of pressure involves the term $\eta^{-1}p^{(-1)}$, which gives rise to pressure gradient $\eta^{-2} g^{(-2)}$ and flux $\eta^{-2}j^{(-2)}$. The balance equation~\eqref{eq:scaledBalanceFlow} then contains term with $\eta^{-3}$ which, when brought back to the $\x$ reference system and multiplied by $\eta^3$, reads
\begin{align}
\label{eq:balance-3}
\eta^{-2}  \sum_{Q\in W} S^{\star} j^{(-2)} &= 0
\end{align} 
Equation~\eqref{eq:balance-3} must be satisfied for all control volumes and periodicity in the $y$ reference system must be obeyed. An~obvious solution and also the only possible one in the case of a~linear behavior is constant pressure $\eta^{-1}p^{-1}$ over the whole RVE resulting in zero pressure gradient and flux $\eta^{-2} g^{(-2)} = \eta^{-2}j^{(-2)} = 0$.  Since the pressure term with the lowest $\eta$ power prescribed by boundary conditions is $p^{(0)}$, the pressure $\eta^{-1}p^{(-1)}=0$ everywhere. There might exist other solutions to Eq.~\eqref{eq:balance-3} in the nonlinear regime of the constitutive relation but one can reasonably assume that the trivial solution is always the correct one.

\subsection{$\eta^{-2}$: constantness of macroscopic fields over RVE \label{sec:rbm}} 
Terms with the negative second power of $\eta$ brought back to the $\x$ reference system and multiplied by $\eta^3$,  $\eta^3$ and  $\eta^4$, respectively, yield 
\begin{subequations} \label{eq:balance-2}
\begin{align}
\eta^{-1}  \sum_{Q\in W} S^{\star} j^{(-1)} &= 0
\\ 
\eta^{-1}  \sum_{J\in V} A^{\star} t^{(-1)}_{\alpha} \e_{\alpha}&=  \bm{0}
\\
\sum_{J\in V} A^{\star}\left(\w^{(0)} + m^{(0)}_{\alpha}\e_{\alpha}\right) &= \bm{0}
\end{align}
\end{subequations} 
Assembling these three equations for all particles and control volumes inside the RVE and considering the periodic boundary conditions, we obtain system with degrees of freedom $p^{(0)}$, $\uu^{(0)}$ and $\eta^{-1}\bomega^{(-1)}$. Let us again consider linearity of the constitutive model. The solution must then be constant pressure $p^{(0)}$ and rigid body motion of the whole RVE defined as $\uu^{(0)} = \vv^{(0)} + \eta^{-1}\llevicivita:\left(\bomega^{(-1)} \otimes \x\right)$ providing zero corresponding strain- and stress-like variables as listed bellow Eq.~\eqref{eq:stressdecomp}. Moreover, the volumetric strain $\varepsilon^{(-1)}_V$, normal crack openings $\delta^{(0)}_N$ and $\delta^{(0)}_{\lambda}$  and relative crack volume $v_c^{(-1)}$ are zero as well because they depend on the strain $\bvarepsilon^{(-1)}$. Contrarily to what was done in Ref.~\parencite{RezCus16}, the solutions of Eqs.~\ref{eq:balance-2} must be periodic in the $y$ coordinate, and therefore one must conclude that $\bomega^{(-1)}=0$. This renders the displacement $\uu^{(0)}$ constant over the RVE. This is the same result one obtains at the same length scale and using the periodicity assumption in the classical mathematical homogenization \parencite{FisShe-97,GhoKyu-01}.

Similarly to the previous scale, other solutions might exist in the nonlinear regime, but the presented trivial solution to system~\eqref{eq:balance-2} will always be assumed to be the only correct one. The variables $p^{(0)}$ and $\uu^{(0)}$ remain unknown. They represent constant macroscopic pressure over the RVE and the RVE macroscopic translation. Thanks to the constantness of $p^{(0)}$, one may replace the averages of the pressure by a~constant hereinafter: $p^{(0)}_{\lambda}=p^{(0)}_{a}=p^{(0)}$.

\subsection{$\eta^{-1}$: RVE problem} 
The negative first power of $\eta$ collects the following terms (already transformed to $\x$ reference system and multiplied by $\eta^3$,  $\eta^3$ and  $\eta^4$, respectively) 
\begin{subequations} \label{eq:balance-1}
%\begin{empheq}[box=\mathbox]{align}
\begin{align}
\sum_{Q\in W} S^{\star} j^{(0)}  &= 0 \label{eq:balance-1flow}
\\
\sum_{J\in V}A^{\star} s^{(0)}_{\alpha} \e_{\alpha} &= p^{(0)} \sum_{J\in V}A^{\star}  b\e_N \label{eq:balance-1forces}
\\
\sum_{J\in V}A^{\star}\left( s^{(0)}_{\alpha} \e_{\alpha} \llevicivita:(\cc_I\otimes \e_N) +  \eta  m^{(1)}_{\alpha}\e_{\alpha}\right) &= p^{(0)} \llevicivita:\sum_{J\in V}  A^{\star}  b \cc_I\otimes \e_N \label{eq:balance-1moments}
%\end{empheq}
\end{align}
\end{subequations}
Note that terms containing  $v^{(-1)}_c$ and $\varepsilon^{(-1)}_V$ are already removed from the flux balance equation because these variables are zero. Also, traction $\bt^{(0)}$ and moment of traction $\eta\w^{(1)}$ are replaced according to their definitions from Eqs.~\eqref{eq:stressSecondTermsTraction} and  \eqref{eq:stressSecondTermsMomentOfTraction}, $\s^{(0)}=f_s(\bvarepsilon^{(0)})$ is the traction in the solid.

The pressure gradients, strains and curvatures that give rise to fluxes, tractions, couple tractions and moments of tractions are $g^{(0)}$, $\bvarepsilon^{(0)}$ and $\eta^{-1}\bchi^{(-1)}$ defined by Eqs.~\eqref{eq:g0}, \eqref{eq:str0} and \eqref{eq:curv-1}. These equations are rewritten based on the previous solution at $\eta^{-2}$ and $\eta^{-3}$ scale ($\eta^{-1}p^{(-1)}=0$, $\eta^{-1}\bomega^{(-1)}=\bm{0}$, $\uu^{(0)}$ and $p^{(0)}$ are constant) and using identity  $\y_{I\! J} = \tilde{\cc}_I-\tilde{\cc}_J$ (Fig.~\ref{fig:2Dsketch}b)
\begin{subequations} \label{eq:RVEstrains}
	\begin{align}
	g^{(0)} =& \frac{\eta}{h}\left[p^{1Q} - p^{1P} \right] - \hat{g} 
	\\
	\varepsilon^{(0)}_{\alpha} =& \frac{e^{\alpha}_i}{l}\left[\eta u^{1J}_i - \eta u^{1I}_i 
	+ \levicivita_{ijk}\omega^{0J}_j c^J_k - \levicivita_{ijk}\omega^{0I}_j c^I_k
	\right] - \hat{\varepsilon}_{\alpha}
	\\
	\eta^{-1}\chi^{(-1)}_{\alpha} =& \frac{e^{\alpha}_i}{l}\left[ \omega^{0J}_i  - \omega^{0I}_i  \right]
	\end{align}
\end{subequations}
where eigen-pressure gradient $\hat{g}$ and eigen-strain $\hat{\bvarepsilon}_{\alpha}$ turn out to be negative projections of the macroscopic pressure gradient $\ba=\nabla_X p^{(0)}$ and the Cosserat strain tensor $\bgamma=\nabla_X\otimes \uu^{(0)} - \llevicivita\cdot\bvarphi^{(0)}$
\begin{subequations} \label{eq:eigenterms}
	\begin{align}
\hat{g} &= -\frac{\partial p^{(0)}}{\partial X_i} e^{\lambda}_i = -\ba \cdot \e_{\lambda} \label{eq:eigen_pgrad}
\\
\hat{\varepsilon}_{\alpha} &= - e^{\alpha}_i \left[ \frac{\partial u^{(0)}_i}{\partial X_k} - \levicivita_{ijk}\varphi^{(0)}_j\right] e^N_k = -\e_N \cdot \bgamma \cdot \e_{\alpha} \label{eq:eigen_strain}
\end{align}
\end{subequations}
Both vector $\ba$ and tensor $\bgamma$ are provided by the macroscopic problem to be described later. Vector $\bvarphi^{(0)}$ shall be identified from its definition as a~macroscopic rotation of the RVE and therefore is constant over the RVE. The projection of the macroscopic Cosserat curvature that was derived in Ref.~\parencite{RezCus16} is missing here because in the current derivation $\bomega^{(-1)}=0$.

All three balance equations~\eqref{eq:balance-1} to be solved numerically are steady state (or static) equations; the transient terms are not present. Unknown fields in these  problems are $\eta p^{(1)}$, $\eta\uu^{(1)}$ and  $\bomega^{(0)}$ being the microscopic pressure, translation and rotation, respectively. Note that these problems are partially decoupled. The mechanical RVE depends only on the macroscopic pressure $p^{(0)}$, primary field $\eta p^{(1)}$ has no effect on the mechanical behavior. The transport problem, however, depends on mechanical primary field $\eta\uu^{(1)}$ via the crack opening $\eta\delta^{(1)}_N$. One should therefore first solve the mechanical RVE problem and then use the computed crack openings $\eta\delta^{(1)}_N$ when solving the transport RVE problem. 

The transport RVE is loaded only by the projection of the macroscopic pressure gradient in the form of eigen-pressure gradient. The load applied to the mechanical RVE comes from the macroscale in two ways: (i) in the form of eigen-strain and (ii) as a~force and moment acting on each particle due to the macroscopic fluid  pressure as the right-hand side terms in Eqs.~\eqref{eq:balance-1}. 

The periodic boundary conditions must be enforced for pressure, displacements and rotations. Furthermore, equations~\eqref{eq:RVEstrains} consider only differences in the primary fields of displacements and pressure and are ill-conditioned without an~additional constraint. The assumption behind the asymptotic expansion~\eqref{eq:expansion} requires the microscopic fields (fluctuations) to be zero on average, therefore the last boundary conditions should prescribe zero volumetric average of these fields over the RVE
\begin{align} \label{eg:RVE_BC}
\langle \eta  p^{(1)}\rangle &= 0 
&
\langle \eta  \uu^{(1)}\rangle &= \bm{0} 
\end{align} 
where the weighted volumetric average reads 
\begin{align}
\langle\bullet\rangle = \frac{1}{\VRVE} \sum_{w\in \VRVE} V_w \bullet \label{eq:volumAver}
\end{align} 
with $V_w$ being volume ($V_I$ or $W_P$) associated with the mechanical or mass transport  node and $w$ denotes either mechanical ($e$) or conduit ($d$) elements.

Direct enforcement of boundary conditions~\eqref{eg:RVE_BC} is not practical. If one applies the linear constraint, the system matrix becomes full and computational and storage requirements rapidly grow. It is therefore recommended to randomly pick some node where pressure and translations are directly prescribed to be some random values (the easiest is to set everything to zero). After the solution is found, both pressure and displacement fast fields can be shifted to satisfy the required constraint, i.e.,~equations~\eqref{eg:RVE_BC} are enforced during the post-processing. Moreover, the actual fast primary fields are typically not required and one can skip this post-processing step.

\subsection{$\eta^0$: macroscopic level} 
Terms with zero power of $\eta$, already transformed to $\x$ reference system and multiplied by $\eta^3$,  $\eta^3$ and  $\eta^4$, respectively, are the following (note that terms with $v_c^{(1)}$ and $\bomega^{(-1)}$, which are always zero, are already deleted) 
\begin{subequations} \label{eq:balance0all}
\begin{align}
\sum_{Q\in W} \left[ \eta S^{\star} j^{(1)}  - \rho_{w0} W \dot{v}^{(0)}_c\left( 1 - b + \frac{p^{(0)}-p_0}{K_{w}} \right)  - \rho_{w0} W v^{(0)}_c \frac{\dot{p}^{(0)}}{K_w}  \right] &= \rho_{w0}\left( 3 b \dot{\varepsilon}^{(0)}_V + \frac{\dot{p}^{(0)}}{M_b}\right) W + Wq  \label{eq:balance0fluxes}
\\
\eta\sum_{J\in V} A^{\star}  t^{(1)}_{\alpha} &= V\rho \ddot{\uu}^{(0)} - V \bb    \label{eq:balance0forces}
\\
\eta^2\sum_{J\in V} A^{\star}\left(\w^{(2)} + m^{(2)}_{\alpha} \e_{\alpha} \right) &= \M_{u\theta}^T \cdot \ddot{\uu}^{(0)} \label{eq:balance0moments}
\end{align}
\end{subequations}

The balance of the whole RVE unit is of interest at the macrolevel, therefore equations for all the bodies within the RVE are summed and divided by the RVE volume, $\VRVE$. One arrives, after several mathematical modifications reported in detail in Appendix~\ref{app:macro}, to partial differential equations describing balance of mass and linear and angular momentum. The momentum balance equations correspond to Cosserat (micromorphic) continuum coupled with mass transport via Biot's theory and cracking
\begin{subequations}\label{eq:macrofinal}
%\begin{empheq}[box=\mathbox]{align}
\begin{align}
\nabla_X\cdot \f &= \rho_{w0}\left[ \dot{v}_{c0}\left(1-b+\frac{p^{(0)}-p_0}{ K_{w}}\right)  +  v_{c0} \frac{\dot{p}^{(0)}}{ K_w} + 3 b \dot{\varepsilon}^{(0)}_V +   \frac{\dot{p}^{(0)}}{ M_b }   \right] + q \label{eq:macrofinaltransport}
\\
\nabla_X\cdot \bsigma_{\mathrm{s}} - \nabla_X p^{(0)} \cdot \bxi  &=  \langle\rho\rangle \ddot{\uu}^{(0)} - \bb \label{eq:macrofinalforces}
\\
\nabla_X\cdot \bmu_{\mathrm{s}} - \nabla_X p^{(0)} \cdot \bzeta + \llevicivita:\bsigma_{\mathrm{s}} - p^{(0)}\llevicivita:\bxi&= \bm{0}\label{eq:macrofinalmoments}
%\end{empheq}
\end{align}
\end{subequations}
with the following macroscopic tensors transforming information from the RVE to the macroscale
\begin{subequations} \label{eq:finalmacrostress}
%\begin{empheq}[box=\mathbox]{align}
\begin{align}
\f &= \frac{1}{\VRVE}\sum_{d\in\VRVE}  hS^{\star} j^{(0)} \e_{\lambda} \label{eq:macroflux}
\\
\bsigma_{\mathrm{s}} &= \frac{1}{\VRVE}\sum_{e\in\VRVE}  lA^{\star} s^{(0)}_{\alpha} \e_{N} \otimes \e_{\alpha} \label{eq:macrostress}
\\ 
\bmu_{\mathrm{s}} &= \frac{1}{\VRVE}\sum_{e\in\VRVE}  lA^{\star} \e_{N} \otimes \left[ \eta m^{(1)}_{\alpha}  \e_{\alpha} + s^{(0)}_{\alpha} \llevicivita : \left( \x_c \otimes \e_{\alpha}\right) \right] \label{eq:macrocouplestress}
\\
\bxi &= \frac{1}{\VRVE}\sum_{e\in\VRVE}  lA^{\star} b \e_{N} \otimes \e_{N} \label{eq:xi}
\\
\bzeta &= \frac{1}{\VRVE}\sum_{e\in\VRVE}  lA^{\star}b \e_{N} \otimes \left[ \llevicivita : \left( \x_c \otimes \e_{N}\right) \right] \label{eq:zeta}
%\end{empheq} 
\end{align}
\end{subequations}
$\f$ is the flux vector, $\bsigma_{\mathrm{s}}$ is the solid stress tensor, $\bmu_{\mathrm{s}}$ is the solid couple stress tensor and $\bxi$ and $\bzeta$ are second order tensors describing RVE internal structure producing an~effect of pressure on the macroscopic stress and couple stress. As the RVE geometry remains unchanged during the calculation, $\bxi$ and $\bzeta$ are constant tensors evaluated only once at the simulation initiation. Evaluation of volumetric strain $\varepsilon^{(0)}_V$ used in Eq.~\eqref{eq:macrofinaltransport} is derived in Appendix~\ref{ap:volumetric_strain_expansion}.

The primary fields (unknowns) at the macroscale are pressures $p^{(0)}$, displacements $\uu^{(0)}$ and rotations $\bvarphi^{(0)}$. Several emerging coupling terms are obtained describing storage of the fluid in cracks and an~effect of volumetric changes on the pressure (Eq.~\ref{eq:macrofinaltransport}), an~effect of the pressure gradient on the linear momentum balance (Eq.~\ref{eq:macrofinalforces}), and effects of the pressure gradient and the pressure on the angular momentum balance (Eq.~\ref{eq:macrofinalmoments}). The macroscopic equation are naturally anisotropic due to the heterogeneity and cracking at the microscale.

The macroscale problem shall be supplemented with appropriate boundary conditions and solved, e.g., by finite element method. At each integration point, the macroscopic pressure gradient $\ba$ and the Cosserat strain tensor $\bgamma$ are computed and projected onto the RVE problem (Eqs.~\ref{eq:eigenterms}) from which stress-like variables (Eqs.~\ref{eq:finalmacrostress}) are evaluated. According to~Ref.~\parencite{RezCus16} the Cosserat effect does not contribute in a~significant way and simplification to Cauchy continuum is possible. Also note that no transient term appears in the balance of angular momentum~\eqref{eq:macrofinalmoments} at this scale.  

\section{Parallel normal and contact vectors \label{sec:paralelism}}

The equations can be further simplified under two assumptions: (i) the Biot coefficient $b$ is constant within the RVE and (ii) the normal and contact vectors are parallel, $\e_N||\n$. The first assumption is satisfied when an~identical material is used in the whole RVE domain and allows to move Biot coefficient in front of the summations. The second assumption holds when model geometry is based on Voronoi or power/Laguerre tessellation and ensures that $A^{\star} = A$.

Let us first derive two useful identities from the divergence theorem. When the area multiplied by outward normal is summed over enclosed surface $\Gamma$, the following holds
\begin{align}
\sum_{J\in V}A\e_N  
= \int_{\Gamma} e^N_i \dd{\Gamma} = \int_{\Gamma} \bi \cdot \e_N \dd{\Gamma} = \int_{V} \nabla\cdot \bi \dd{V}  =\bm{0}  \label{eq:identityA}
\end{align}
The second identity sums multiplication of area, its centroid and outward normal over enclosed surface of some discrete body around node $I$
\begin{align}
\sum_{J\in V} A \cc_I\otimes \e_N = \int_{\Gamma} \rr \otimes \e_N \dd{\Gamma} = 
\int_{\Gamma} r_i e^N_j \dd{\Gamma} = \int_{\Gamma} r_i \bj \cdot \e_N \dd{\Gamma} = \int_{V} \nabla \cdot (r_i \bj)\dd{V}  = \int_{V} \frac{\partial r_i}{\partial X_j}\dd{V} = \int_{V} \delta_{ij}\dd{V} = \bm{1} V  \label{eq:identityB}
\end{align} 
The vector $\rr$ serves as a~general position vector pointing from node $\x_I$ to any point on the particle surface or in its volume, $\bi$ and $\bj$ are unit Cartesian basis vectors and $\bm{1}=\delta_{ij}$ is the second order identity tensor (Kronecker delta).

Let us first show that right-hand sides of Eqs.~\eqref{eq:balance-1forces} and \eqref{eq:balance-1moments} disappear. Simple application of identities \eqref{eq:identityA} and \eqref{eq:identityB} provides
\begin{subequations}
\begin{align}
p^{(0)} b \sum_{J\in V}A \e_N &= 0\\
p^{(0)} b \llevicivita:\sum_{J\in V}  A \cc_I\otimes \e_N  = p^{(0)} b V \llevicivita : \bm{1}   &= \bm{0}
\end{align}
\end{subequations}
Therefore, constant Biot coefficient and parallelism $\e_N||\n$ make the mechanical RVE problem completely independent of the pressure. 

Moreover, tensor $\bxi$ from Eq.~\eqref{eq:xi} can be simplified.  Recalling $l\e_N = \x^{I\!J} =\cc_I-\cc_J$, the summations can be rewritten from summing over mechanical elements to double summations over rigid bodies and then over all its neighbors 
\begin{align}
\bxi &= \frac{b}{\VRVE}\sum_{e\in\VRVE}  A (\cc_I-\cc_J) \otimes \e_{N} = \frac{b}{\VRVE}\sum_{I\in\VRVE}\sum_{J\in V}  A \cc_I \otimes \e_{N} = \frac{b}{\VRVE}\sum_{I\in\VRVE} V \bm{1} = b\bm{1}
\end{align} 
Therefore, in Eq.~\eqref{eq:macrofinalmoments}, the term with $\bxi$ cancels out as $\llevicivita:\bm{1}=\bm{0}$.

Finally, let us also comment on tensor $\bzeta$. The assumption of constant Biot coefficient $b$ and parallel normal and contact vector does not simplify it. This tensor provides, when multiplied by pressure gradient, overall moment density due to the pressure gradient. It would be zero only for very specific case of no eccentricities -- when all vectors $\cc_I$ and $\cc_J$ are scalar multiple of corresponding normal vector $\e_N$. This implies that the governing nodes bearing degrees of freedom are located at centroids of the rigid bodies. The described situation is theoretically achievable for some regular arrangements or when using centroidal Voronoi tessellation. However, both of these cases introduce large directional bias into the model and are of no practical use.

\section{Implementation}
The \emph{full} and \emph{homogenized} models are implemented in an~in-house software. Both steady state and transient simulations are presented in the following text, however, the transient terms are only used for the mass transport part; mechanical behavior is always quasi-static in the following verification studies. Implicit time integration scheme called the generalized-$\alpha$ method with spectral radius 0.8 is adopted~\parencite{CarKuh12,JanWhi-00}; the two-way coupled problem of mechanical equilibrium and transport is solved in a~strongly coupled numerical scheme ensuring second order accuracy. 

The mechanical constitutive model (function $f_s$ from Eq.~\ref{eq:LDPMconst}) is a~simplified older version of the Lattice-Discrete Particle Model (LDPM) according to Ref.~\parencite{CusCed07}. The simplification consists mostly in reducing the number of material parameters to 4 and omitting the confinement effect. Nevertheless, the full LDPM constitutive model can be easily used in the homogenization. The function $f_m$ from Eq.~\eqref{eq:LDPMconst} relating curvature and couple traction at the contact face is assumed zero.  The transport constitutive relation (function $f_j$ from Eq.~\ref{eq:LDPMconst}) is a~simple linear dependency of flux on the pressure gradient in intact material with additional terms according to Ref.~\parencite{GraBol16} when cracks are present. The nonlinear behavior in uncracked material can also be easily homogenized as is done in Ref.~\parencite{EliYin-22}. Both mechanical and transport constitutive formulations are described in detail in Appendix~\ref{app:A}, which includes values of all the material model parameters listed in Tab.~\ref{tab:materparams}. 

The spatial discretization is based on an~actual mesostructure, it shall be called \emph{physical} according to Ref.~\parencite{BolEli-21}. Radii of the spherical particles are obtained from the Fuller curve~\parencite{FulTho07}. The location of the particles in the domain is generated randomly in a~sequence restricting overlapping. The power/Laguerre tessellation provides transport connectivity and the discrete bodies for mechanics while (weighted) Delaunay triangulation gives mechanical connectivity and control volumes for the transport part. Such discretization ensures parallelism of normal and contact vectors: $\oo||\e_{\lambda}$ and $\e_N||\n$, therefore $S\equiv S^{\star}$ and $A\equiv A^{\star}$. Since Biot coefficient is kept spatially constant, the simplifications described in Sec.~\ref{sec:paralelism} hold. All the verification examples assume material with maximum aggregate diameter $d_{\max}=10$\,mm and aggregate relative volumetric content 80\%. For sake of computational feasibility, only aggregates with diameter above 4\,mm are explicitly considered, the rest is phenomenologically represented by the contact constitutive behavior.  

\textcite{RezCus16} showed that cubic RVE with edge length $5d_{\max}$ is already sufficient for the mechanical problem. According to Ref.~\parencite{EliYin-22}, such RVE size is acceptable also for the transport problem. Therefore, the RVE used here for all the calculation has size $50\times50\times50\,$mm$^3$; the mechanical RVE has 1,539 degrees of freedom (DoF) and the transport RVE has 2,160\,DoF. A~nonlinear steady-state solver is used to calculate the RVE problems.

In the initial intact state, responses of both mechanical and transport RVEs are linear. They can be therefore easily pre-computed resulting in a~great computational cost reduction. If one is interested in macroscopic variables only, there is no need to reconstruct the fast field and pre-computed material matrix is sufficient. The linear pre-computed state is adaptively replaced by the full RVE non-linear problem with a~help of Ottosen's criterion~\parencite{Ott77} serving as an~indicator of inelasticity. The decision process is implemented exactly as described in Ref.~\parencite{RezZho-17} for the mechanical homogenization, the transport RVE is replaced simultaneously. 

The continuous macroscopic solution is approximated via the finite element method, development of the discretized weak form of Eqs.~\eqref{eq:macrofinal} is straightforward. Cosserat trilinear isoparametric brick elements extended by additional pressure degrees of freedom are used. The same trilinear shape functions are applied to approximate the element shape as well as the displacements, rotations and the pressure.  The methodology for element implementation was adopted from Ref.~\parencite{HorPat-14}. Full Gauss integration using 8 integration points (hence 8 submodel RVE pairs for each element) is employed. \textcite{RezCus16} shows that the Cosserat components might be ignored with almost no price in accuracy, however such simplification is not implemented here. 

\section{Verification}

The model is verified by four examples featuring concrete specimens: (i) Terzaghi's consolidation, (ii) flow through a~compressed cylinder, (iii) constrained tension of a~sealed prism and (iv) hydraulic fracturing of a~hollow cylinder.  The examples specifically target different coupling terms in the macroscopic balance equations. All verification examples compare the \emph{full} discrete model and the \emph{homogenized} model. All four cases also share the same material mesostructure (described in the previous section) and constitutive model described in Appendix~\ref{app:A}, including the material parameters from Table~\ref{tab:materparams}. 

The mesoscale discrete model exhibits intrinsic randomness due to random locations of aggregates. Even though the comparison should ideally be repeated several times for different realizations of concrete mesostructure, only single realization is used in all the examples. All the calculations are done on a~single processor within a~single thread. There is a~huge potential to speed-up the \emph{homogenized} model by distributing independent RVE problems over several processors. Such distributed solution is possible also for the \emph{full} model by parallelizing solver of the system of linear equations, but efficiency is typically significantly lower.

\subsection{Linear Terzaghi's consolidation \label{sec:consolidation}}
The model is first verified by simulating Terzaghi's consolidation. A~prism of material of size $0.5\times 0.1\times 0.1$\,m$^3$ is initially under zero pressure and zeros strain. The $x$ axis runs along its longest central axis, the domain begins at $x=0$\,m and ends at $x=0.5$\,m. The prism is sealed at all boundaries except the front end at $x=0$ where pressure $p^{\star}$ is prescribed at time $t=0$ and kept constant throughout the simulation. The mechanical boundary conditions prescribe zero rotations at all boundaries, zero $x$ displacement at the rear end at $x=0.5$\,m and zero $y$ and $z$ displacements at all rectangular sides. Total traction $t_x^{\star}$ is prescribed at the front end at $x=0$ and kept constant the whole simulation time. Biot coefficient $b=0.5$ and reference pressure $p_0=0$\,MPa; linear elastic material behavior is enforced. 

\begin{figure}[tb!]
	\centering\includegraphics[width=12cm]{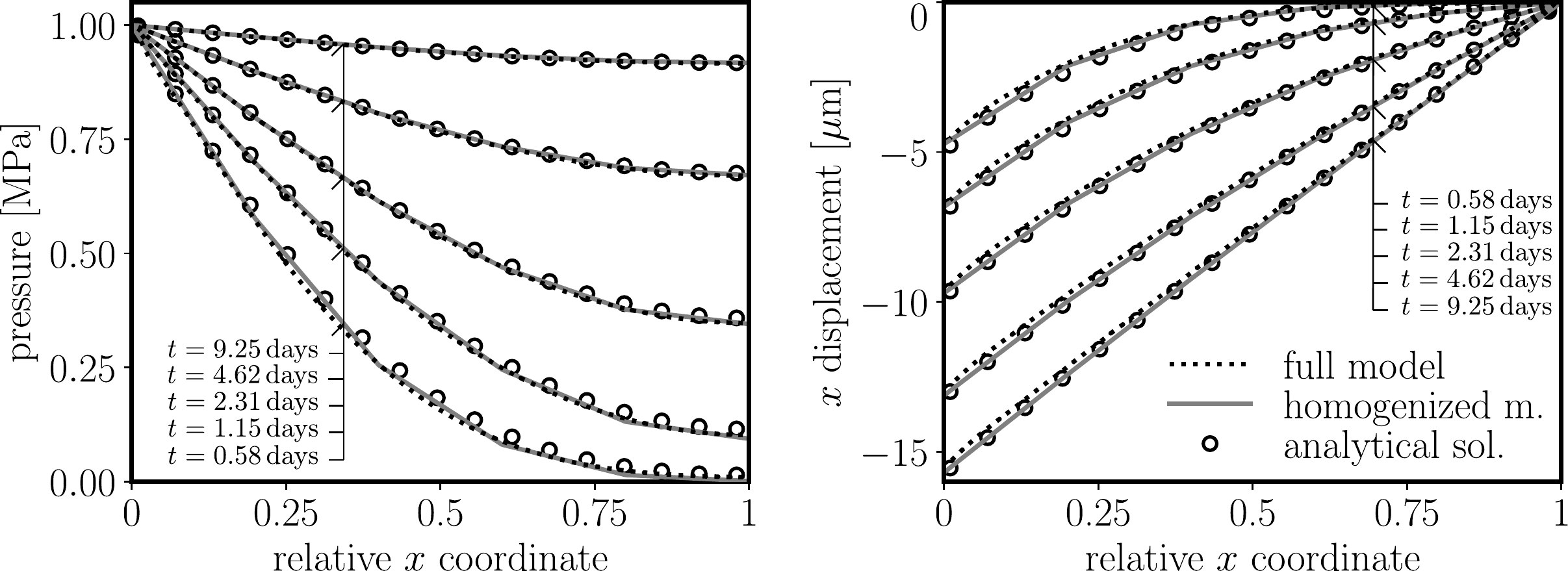}
	\caption{Terzaghi's consolidation, loading by fluid pressure: fluid pressure and $x$ displacement profiles at various times of simulation as computed by the \emph{homogenized} model (macroscopic component only) and the \emph{full} model. }\label{fig:consolidation_pressure}
\end{figure}

\begin{figure}[tb!]
	\centering\includegraphics[width=12cm]{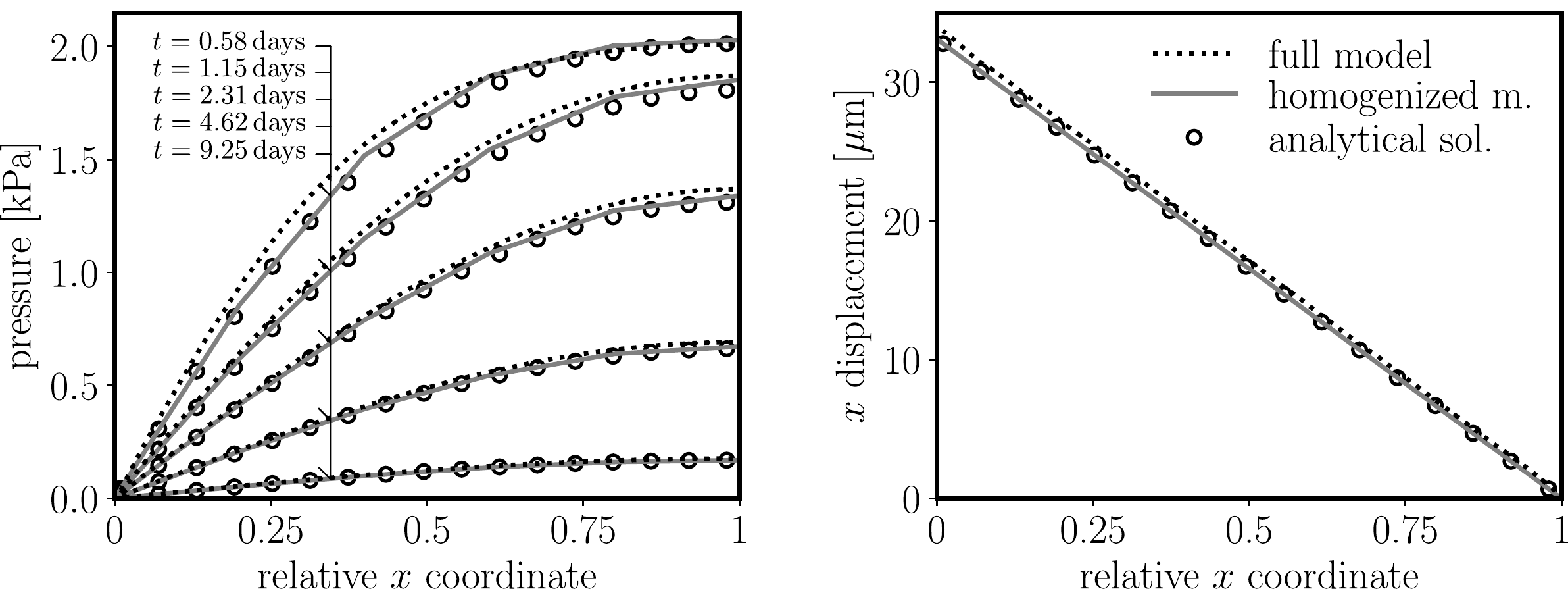}
	\caption{Terzaghi's consolidation, loading by traction: fluid pressure and $x$ displacement profiles at various times of simulation as computed by the \emph{homogenized} model (macroscopic component only) and the \emph{full} model.}\label{fig:consolidation_traction}
\end{figure}

Two cases are investigated. In the first case (denoted \emph{loading by pressure}), the prescribed values read $p^{\star}=1$\,MPa and $t_x^{\star}=0$\,MPa. The pressure propagates along the $x$ axis and transfers into mechanical forces through the Biot's coupling leading to elongation of the prism. Prescribed values for the latter case  (denoted \emph{loading by traction}) read $p^{\star}=0$\,MPa and $t_x^{\star}=1$\,MPa. In this case the mechanical volumetric deformation induces a~pressure increase that is released at the front end of the prism.  

The problem is analyzed using the \emph{full} discrete and \emph{homogenized} continuous model. Five trilinear 8-node isoparametric finite elements of equal size along the $x$ direction are used at macroscale. Comparison of both models in terms of pressure and $x$ displacement profiles along the central $x$ axis at different time instants is presented in Figs.~\ref{fig:consolidation_pressure} (loading by pressure) and \ref{fig:consolidation_traction} (loading by traction). Relatively small differences are attributed to (i) heterogeneity and (ii) boundary layer effect~\parencite{Eli17}, or wall effect, that is present in the \emph{full} model and modify the mechanical properties in the vicinity of the boundary.

The \emph{full} model has approx $124,000$ DoF and the computation took approximately 3\,hours for pressure loading and 2.5\,hours for traction loading case. On contrary, the \emph{homogenized} model has only 40 DoF and calculation took approximately 1\,s in both loading cases. The speed-up factor is about four orders of magnitude thanks to the enforced linearity of the RVE response, which can be pre-computed at the simulation initialization.

There exists~an analytical solution to the problem~\parencite{Ter23,DetChe93}. Pressure and displacement fields are provided by the functions listed in Appendix~\ref{app:B}.  The analytical solution, added to Figs.~\ref{fig:consolidation_pressure} and \ref{fig:consolidation_traction}, agrees well with the numerical model results.

\subsection{Flow through compressed cylinder}
The second verification example is steady state simulation of cylinder (depth 0.1\,m, diameter 0.1\,m) compressed in $z$ direction assuming no friction taking place at the loading platens, i.e.,~the lateral expansion is completely unconstrained. Simultaneously, a~pressure gradient is applied in $z$ direction by prescribing pressure at the bottom face at $z=0$\,m being  1\,MPa and at the top face at $z=0.1$\,m being 0\,MPa. The normal flux component over the cylindrical shell is set to zero. The Biot coefficient is $b=0$, hence the only coupling mechanism is the dependency of the permeability coefficient $\lambda$ on normal crack opening.

\begin{figure}
	\centering\includegraphics[width=10cm]{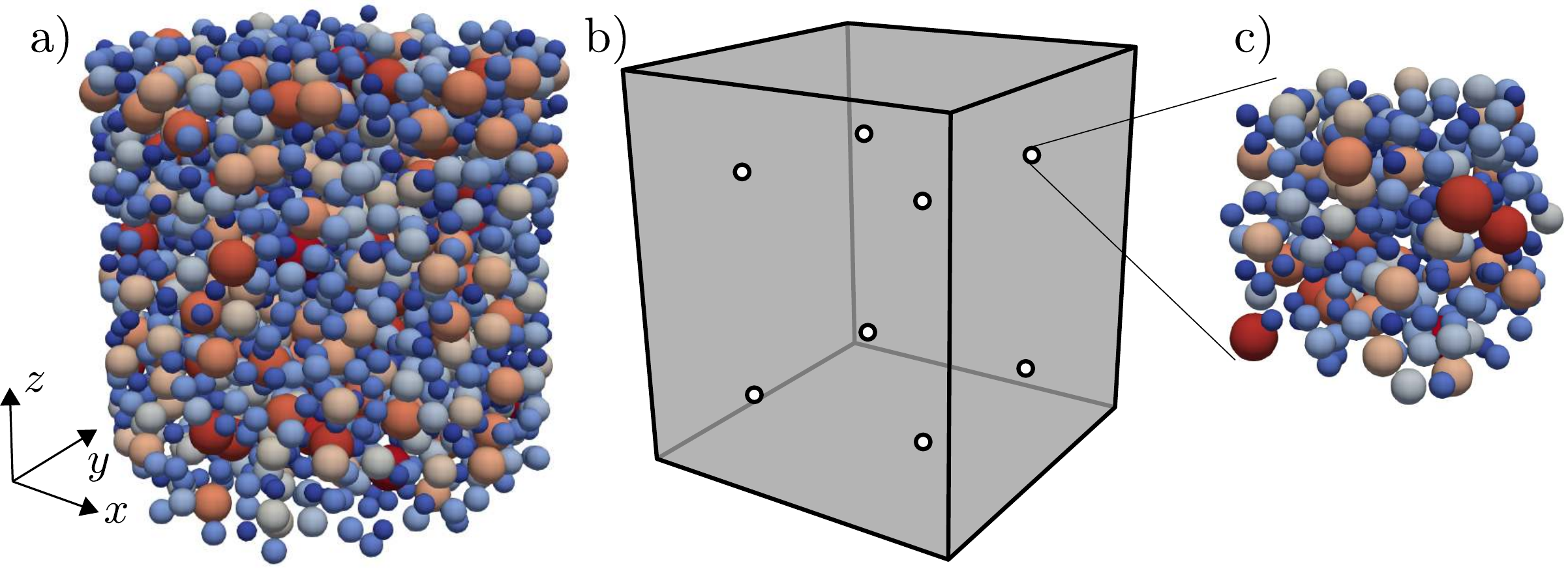}
	\caption{a) Cylindrical specimen, \emph{full} model; b) single brick element, \emph{homogenized} model; c) RVE attached to an~integration point.}\label{fig:compression}
\end{figure}

\begin{figure}
	\centering\includegraphics[width=12cm]{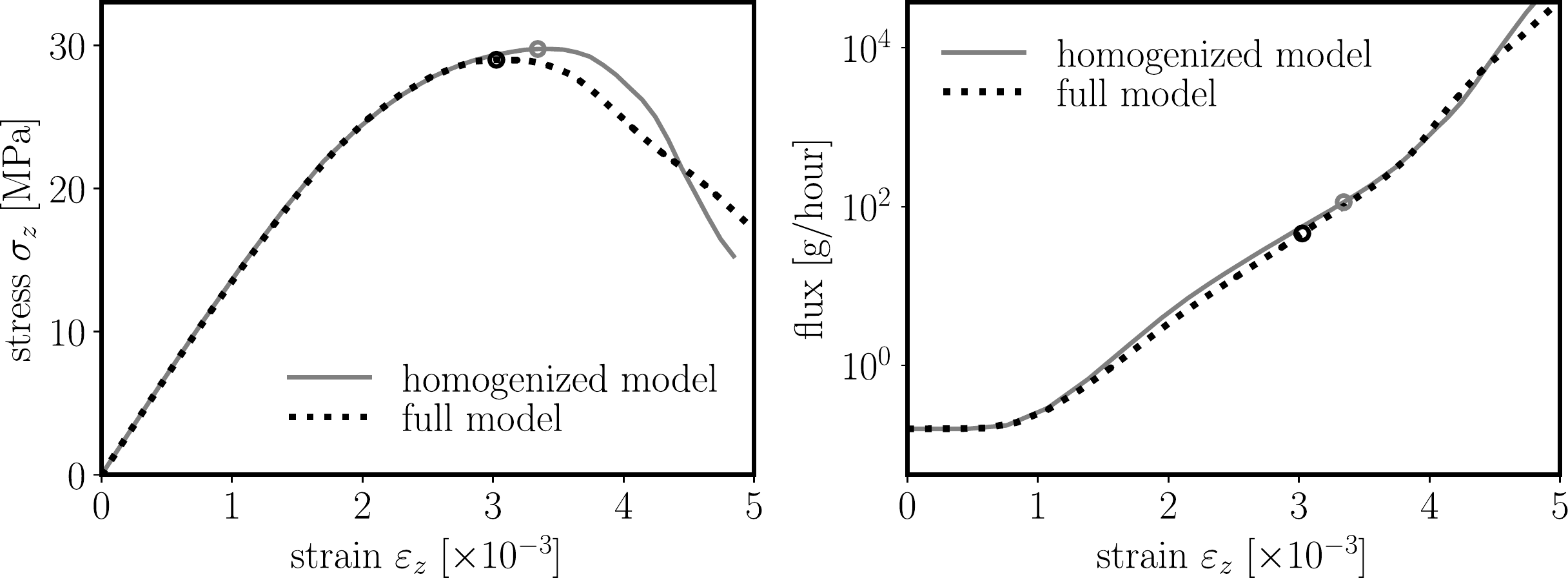}
	\caption{Loading traction and flux through cylindrical specimen obtained by the \emph{full} and \emph{homogenized} model. Circles label states at which the models reached the maximum loading force.
	}\label{fig:res_compression}
\end{figure}

The \emph{homogenized} model consists of a~single brick with 8 integration points. Depth of the brick is 0.1\,m, upper and bottom faces are squares of size $0.05\sqrt{\pi}\times0.05\sqrt{\pi}$\,m$^2$ corresponding to the cylinder cross-sectional area. Each of the 8 RVE pairs is independent, initially pre-computed but later, after the Ottosen's criterion indicates it, represented by the full RVE problem. A~sketch of the model is shown in Fig.~\ref{fig:compression}. The macroscopic model consists of additional 8 free DoF. The \emph{homogenized} model therefore has $8\times(2,160+2,016)+8=33,416$ DoF in total and runs in 25\,min.

The \emph{full} model has $32,500$\,DoF and runs in 68\,min. Even though the number of DoF is similar, the speed-up factor is about 2.7. The improvement is caused by decoupling the systems into separate problems, hence speeding-up its solution.

Stresses and fluxes obtained by the \emph{full} and \emph{homogenized} model are similar, see Fig.~\ref{fig:res_compression}. Since the constitutive model features strain softening, the mechanical problem suffers from strain localization. It is well known that the localization phenomenon cannot be homogenized as the scale separation does not hold~\parencite{GitAsk-07}. Few research papers~\parencite{CoeKou-12b,Ung13} provide suggestions and remedies to capture strain localization properly in the homogenization scheme (see Sec.~\ref{sec:intro}) but these are not used here. The reason why our verification study gives more or less satisfactory results even under strain localization is that the RVE volumes exactly correspond to the volumes of material represented by the associated integration points. Consequently, the strain localization  occurs in the same material volumes in both models and should be therefore macroscopically equivalent. This shortcut was developed in Ref.~\parencite{RezCus16} for the mechanical problem only, one can see that it is applicable also for the coupled problem. Unfortunately, the major benefit of homogenization, that the large material volume can be macroscopically represented by small RVE, is lost.

\subsection{Constrained tension of a~sealed prism}

The third verification example is focused on coupling terms excluded in the previous two examples, specifically the effect of cracks on fluid stored in the transient simulation. The same prism as in the consolidation example~\ref{sec:consolidation} is used. To completely exclude complications with strain localization, the mechanical fields (displacement and rotations) are strongly constrained: three macroscopic degrees of freedom with meaning of macroscopic strains $\varepsilon_x$, $\varepsilon_y$ and $\varepsilon_z$ dictate displacement of all the particles while all the rotations are zero.
\begin{align}
u_i &= x_i \varepsilon_i & \theta_i &= 0 & i\in&\{x,\,y,\,z\}\label{eq:Voigt}
\end{align}
The strain $\varepsilon_x$ is prescribed, growing by rate $2\times10^{-8}$\,s$^{-1}$, the other two strains are free. Stress $\sigma_x$ is measured as conjugate variable to $\varepsilon_x$ divided by the prism volume. The whole mechanical problem has only 2 degrees of freedom ($\varepsilon_y$ and $\varepsilon_z$) under such constraint. The transport problem is unconstrained and sealed, no flux is allowed between the prism domain and the surroundings. The initial pressure as well as the reference pressure $p_0$ are set to $1$\,MPa and the mechanical DoF $\varepsilon_y$ and $\varepsilon_z$ are initially set so that the model is in static equilibrium. Due to applied mechanical tension, the average pressure in the domain decreases and the prism shrinks in lateral direction. Simulations are performed with various values of the Biot coefficient $b$ ranging from 0 to 1. 

\begin{figure}
	\centering\includegraphics[width=\textwidth]{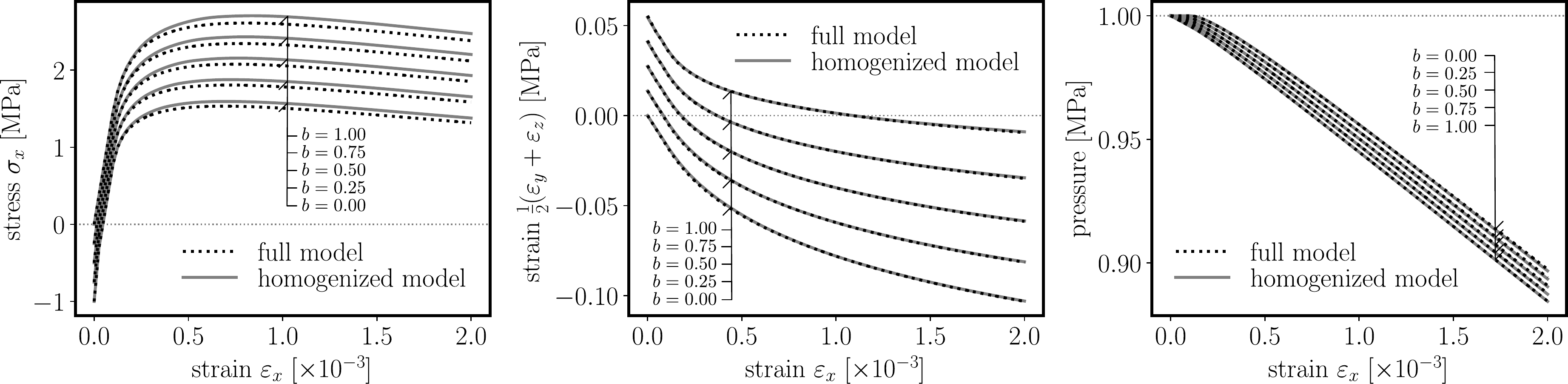}
	\caption{Evolution of stress $\sigma_x$, the average of lateral macroscopic strains and the average pressure in prism loaded by strain $\varepsilon_x$ under Voigt's constraint. }\label{fig:pres_mech}
\end{figure}

The Voigt's constraint described by Eq.~\eqref{eq:Voigt} is applied also to the \emph{homogenized} model at macroscale. The microscopic mechanical field must be zero everywhere, therefore all displacements and rotations at mechanical RVEs are zero (no free DoF is left). Note that the RVE cracks still develop due to the eigen-strains projected from the macroscale. Around $83,000$\,DoF are used for the transport part in the \emph{full} model. 24 macroscopic pressure DoF are free in the \emph{homogenized} model, summing with 40 transport RVEs of $2,160$\,DoF makes it $24 + 40\times2,160=86,424$ transport DoF in total. The average computation times for \emph{full} or \emph{homogenized} models are around 3\,hours or 40\,min, respectively.  Speed-up factor 4.5 is again achieved by decoupling the system into individual RVE problems, even though the total number of DoF is comparable. 

Figure~\ref{fig:pres_mech} presents obtained axial macroscopic stress, lateral strain and the average pressure. There is an~excellent agreement between the \emph{full} and \emph{homogenized} models, minor difference is seen in stress $\sigma_x$. This  is again attributed to the wall effect present in the \emph{full} model~\parencite{Eli17}. Note, that the lateral strains vary with Biot coefficient; apparent Poisson's ratio is strongly affected by the pressure of the fluid.

\subsection{Hydraulic fracturing of hollow cylinder}

The last example is meant as a~simple application. The specimen is a~hollow cylinder of depth 0.05\,m, inner radius $r_{\mathrm{i}}=0.05$\,m, outer radius $r_{\mathrm{o}}=0.3$\,m. The cylinder central axis is align with axis $z$. Transport boundary conditions prescribe constant zero pressure at the outer cylinder shell, pressure linearly increasing in time $p_{\mathrm{i}}=200t$ at the inner surface and zero normal flux at the upper and bottom surfaces at $z=0$\,m and $z=0.05$\,m. Mechanical boundary conditions restrict $z$ displacements at the upper and bottom surface and apply inward traction on the inner surface of a~magnitude equal to the prescribed pressure. 

The \emph{full} model has approximately $323,000$\,DoF. The \emph{homogenized} model was built in two versions. (i) The fine version is discretized into 15 elements equidistantly distributed in radial direction, 30 elements along the circumference and one element over the depth, resulting in 450 elements and 3600 RVEs. (ii) The coarse version has only 4 elements in radial direction and 10 element along circumference. In total, it has 40 element, 320 RVEs and approx. $320\times(2,160+2,016)\approx1.3$\,mil.\,DoF, assuming all the RVEs are switched from pre-computed to inelastic states.

\begin{figure}[tb]
	\centering\includegraphics[width=\textwidth]{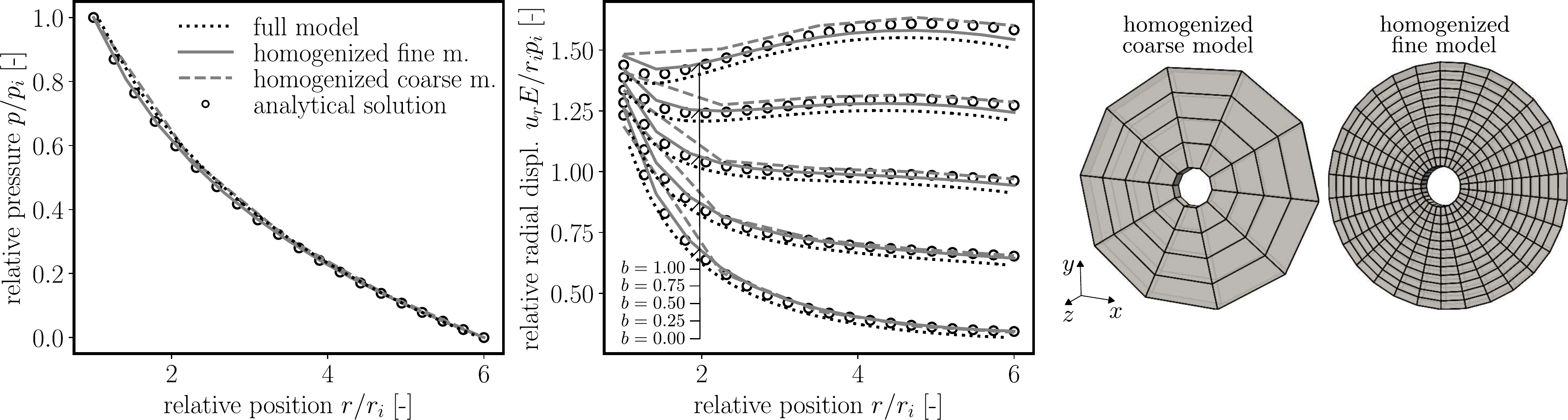}
	\caption{Pressure and radial displacement profiles from the elastic simulations of the pressurized hollow cylinder. The analytical solution is taken from Ref.~\parencite{GraFah-15}. Meshes used for the \emph{homogenized} coarse and fine models are sketched on the right-hand side.}\label{fig:tube_elastic}
\end{figure}

 \begin{figure}[tb]
	\centering\includegraphics[width=\textwidth]{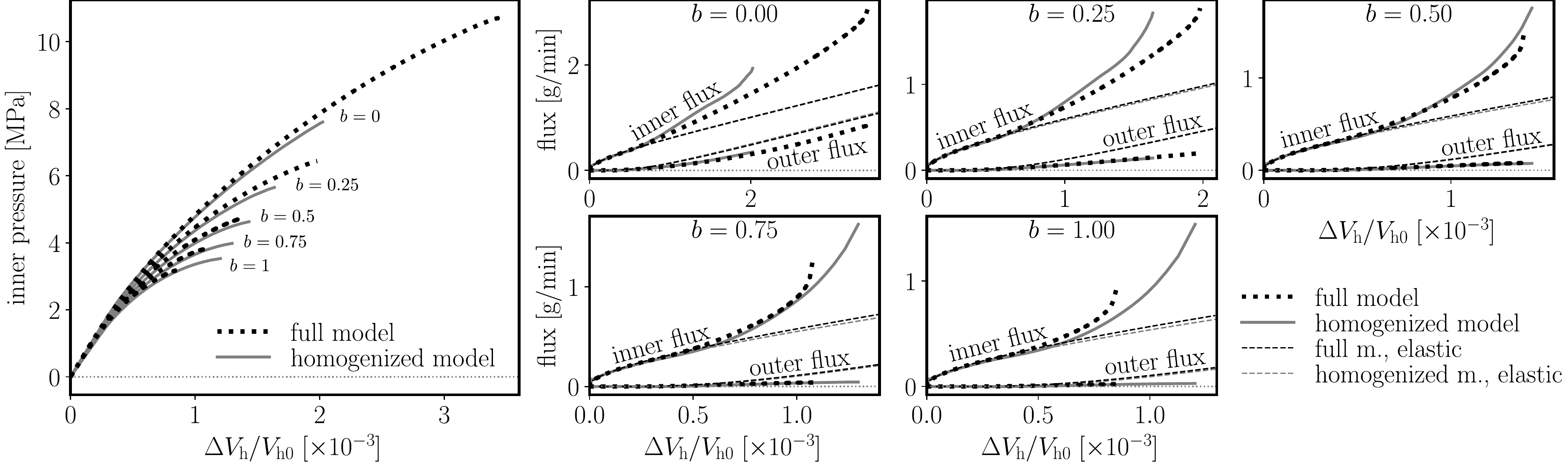}
	\caption{Pressure in the hole and inner and outer fluxes obtained during simulation of hydraulic fracturing. The horizontal axis shows relative volumetric change of the central hole of initial volume $V_{\mathrm{h0}}$, $\Delta V_{\mathrm{h}}$ denotes absolute volume change.}\label{fig:tube_inelastic}
\end{figure}

Let us first verify that the model provides the correct steady state elastic solution. Figure~\ref{fig:tube_elastic} shows relative pressure and displacement profiles along the radial direction, $r$ is a~distance from the central axis of the cylinder. Both fine and coarse \emph{homogenized} models are used, their results are obtain in almost no time (about 5\,s and 3\,s, respectively) because RVE responses can be pre-computed. The \emph{full} model computation takes about 13\,min, speed up factor is 2--3 orders of magnitude. Results of the coarse model suffer from poor macroscopic approximation, but even these seems relatively close to the \emph{full} model data. 

The analytical solution to the elastic problem is derived in Ref.~\parencite{GraFah-15}, see Appendix~\ref{app:C}. It is shown in Fig.~\ref{fig:tube_elastic} along with the numerical results. The agreement is good for all considered Biot coefficient values.

The inelastic transient behavior is simulated next. Initial and reference pressures are set to zero. Simulation is controlled by increase of the inner pressure, $p_{\mathrm{i}}$; therefore it eventually reaches the critical maximum pressure at which macroscopic cracks develop across the whole cylinder and solver fails due to loss of convergence. Another complication arises in \emph{homogenized} models because the macroscopic crack may develop in any direction. Standard periodic boundary conditions cannot handle such situation~\parencite{StrJir11} and advanced approach is needed~\parencite{CoeKou-12a}. Instead, the RVEs are rotated in this study so they are all aligned with the radial direction under which the crack is expected. Such a~remedy is acceptable for this particular example but cannot be used for a~general case where the crack direction is unknown.

 \begin{figure}
	\centering\includegraphics[width=\textwidth]{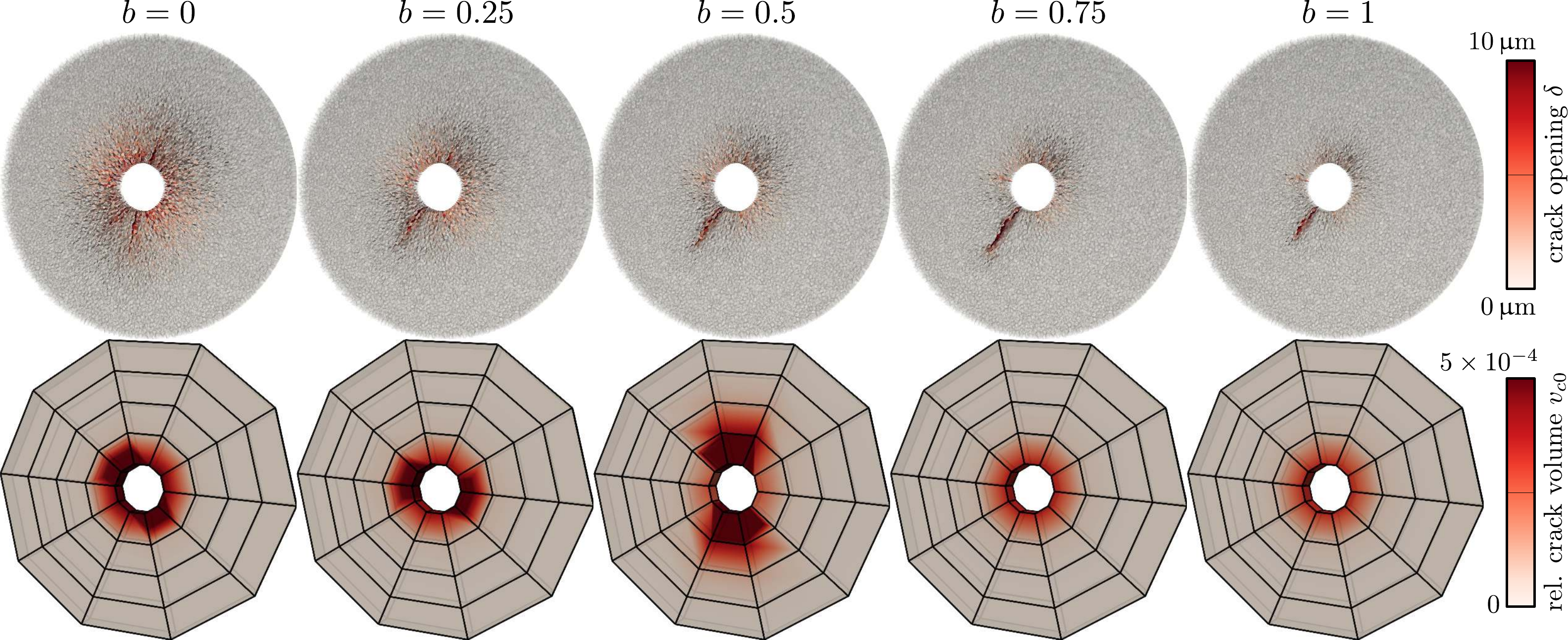}
	\caption{Cracks developed in the \emph{full} and \emph{homogenized} models at the last valid step of the simulation.}\label{fig:tube_cracks}
\end{figure}

Change of the volume of the central hole, $\Delta V_\mathrm{h}$, inner flux and outer flux are recorder during the simulation. The results are shown in Fig.~\ref{fig:tube_inelastic}. A~reasonable correspondence in terms of pressures/tractions and fluxes is obtained. The larger differences occurring in the later stages of the simulations are attributed to the developed strain localization. The RVE volumes of the \emph{homogenized} model are actually much larger than the macroscopic material volumes associated with integration points. Unfortunately, the localized cracks significantly affect the solution and cause deviations of the homogenized model results from the reference \emph{full} solution. Note that Fig.~\ref{fig:tube_inelastic} shows also fluxes obtained from elastic simulations to visualize the effect of cracking on the flux as a~difference between elastic and inelastic solutions.

Figure~\ref{fig:tube_cracks} shows cracks developed in the \emph{full} and \emph{homogenized} models. One can see localized macrocracks developed in the \emph{full} model. The internal structure of the $\emph{full}$ model (positions of particles) is constant, that is why the macrocracks tend to appear at the same locations, presumably at some locally weaker region. On the contrary, the \emph{homogenized} model possesses an~ideal symmetry and strain localization at the macroscale occurs in a~random direction due to the numerical truncation. No strain localization has been achieved for \emph{homogenized} models with Biot's coefficients 0.75 and 1, the symmetry was maintained until the last simulation step. The symmetry can be effectively impaired by either generating several different RVE mesostructures or randomly changing material parameters of otherwise identical RVEs.

Interestingly enough, the cracking character changes with Biot coefficient. There is more diffused microcracking appearing at the critical pressure around the central hole for low Biot coefficients. This is due to different pressure magnitude sustained by models with different Biot coefficient. If one studies the cracking at the same pressure level, the situation is actually the opposite: the larger Biot coefficients give more diffused cracking because the fluid pressure reduces both radial and circumferential compressive stresses in the solid. Since it further helps to open the cracks, the critical pressure for larger Biot coefficients is substantially lower.  

The \emph{full} model simulations took 2970\,min, 2290\,min, 1220\,min, 980\,min and 730\,min for Biot coefficients 0, 0.25, 0.5, 0.75 and 1, respectively. Only the coarse \emph{homogenized} model was employed in the inelastic analyses reaching computational times 330\,min, 250\,min, 150\,min, 132\,min and 120\,min and speed-up factors about 9, 9.2, 8.1, 7.4 and 6.1, respectively. This is surprising considering that the \emph{homogenized} model has about 4 times more DoF when all RVEs leave the pre-computed state. The high speed-up factors are achieved by decoupling the system of equations into independent subsystems and also by keeping many RVEs in their pre-computed state for most of the simulation time.

\section{Conclusions}
Asymptotic expansion homogenization of discrete mesoscale models for coupled mechanics and mass transport in a~fully saturated heterogeneous medium is developed. Coupling is provided by Biot's theory and an~effect of cracks on material permeability coefficient. Based on the results obtained in this study the following conclusions can be drawn.
\begin{itemize}
	\item The macroscale problem arising from the homogenization is described by the standard transient transport and Cosserat continuous differential equations enriched by coupling terms. The macroscopic problem can be solved with a~help of the finite element method by transforming it into the weak form. The constitutive equation routines are replaced by subscale RVE problems. The homogenized macroscopic material exhibits naturally anisotropy due to cracking at the microscale, both for the mass transport and mechanical behavior.
	
	\item The microscale problems are discrete and steady state; the mechanical RVE is independent on the transport RVE while the transport RVE requires information about crack openings from the mechanical RVE. Therefore, they can be solved sequentially. 
	\item Microscale periodic RVE problems are loaded by projection of macroscopic strain and pressure gradient tensors into eigen-strains and eigen-pressure gradients of the discrete elements. The homogenization procedure derives expressions for macroscopic stress, couple stress and flux as a~summation of certain quantities from the discrete elements. These summation formulas correspond to the known expressions used for discrete systems in the literature.
	\item Homogenization techniques derived earlier for pure momentum balance~\parencite{RezCus16} or pure mass transport~\parencite{EliYin-22} emerge from the homogenized coupled system when the coupling terms are removed.
	\item Simplifications are possible for a~special type of tessellation where normal and contact vectors are parallel, e.g., for Voronoi or power/Laguerre tessellations. Several terms at macro and microscale disappear or become simpler. 
	\item The homogenization technique is applied to a~discrete mesoscale model of concrete. Several examples specifically targeting different terms of the coupling structure are provided demonstrating  correspondence between the \emph{full} and \emph{homogenized} model responses. An~additional verification is supplied by comparing the result to the analytical solution, if available. The last example verifying the homogenized equations features hydraulic fracturing of a~hollow cylinder.
	\item Speed-up factors of several orders of magnitude can be obtained for elastic problems for which the RVE responses can be pre-computed. Significant time savings can also be achieved for inelastic simulations. All the comparisons were computed as single thread processes.  It is expected that further improvement would be gained, should they run distributed on several computational cores in parallel because the RVE problems from different integration points can be solved independently. 
	\item The major unsolved problem is strain localization because it violates the fundamental homogenization assumptions and the RVE ceases to exist. The microscale problem becomes size dependent. Recent developments showing remedies to overcome this issue are discussed but not implemented in this study. This issue remains open for further investigation. 
	
\end{itemize}

\newpage
\appendix

\section{Expansion of pressure gradient, strain and curvature \label{ap:strain_expansion}}
The following equations provide individual terms of expansions of the pressure gradient $g$, strain $\bvarepsilon$ and curvature $\bchi$ of the discrete model used in Eq.~\eqref{eq:straindecomp}.
\begin{subequations} \label{eq:gechi_expansion}
	\begin{align}
	g^{(-2)} =&\frac{1}{\tilde{h}}\left[p^{-1Q} - p^{-1P}\right] \label{eq:g-2}
	\\
	g^{(-1)} =&\frac{1}{\tilde{h}}\left[p^{0Q} + \frac{\partial p^{-1Q}}{\partial X_i} y^{PQ}_i - p^{0P}\right] \label{eq:g-1}
	\\
	g^{(0)} =&\frac{1}{\tilde{h}}\left[p^{1Q} + \frac{\partial p^{0Q}}{\partial X_i} y^{PQ}_i  + \frac{1}{2} \frac{\partial^2 p^{-1Q}}{\partial X_i X_j} y^{PQ}_i  y^{PQ}_j - p^{1P}\right] \label{eq:g0}
	\\
	g^{(1)} =& \frac{1}{\tilde{h}}\left[\frac{\partial p^{1Q}}{\partial X_i} y^{PQ}_i + \frac{1}{2} \frac{\partial^2 p^{0Q}}{\partial X_i X_j} y^{PQ}_i  y^{PQ}_j + \frac{1}{6} \frac{\partial^3 p^{-1Q}}{\partial X_i X_j X_k} y^{PQ}_i  y^{PQ}_j y^{PQ}_k\right] \label{eq:g1}
	\\
	\varepsilon^{(-1)}_{\alpha} =&\frac{e^{\alpha}_i}{\tilde{l}}\left[u^{0J}_i - u^{0I}_i
	+\levicivita_{ijk} \omega^{-1J}_j \tilde{c}^J_k - \levicivita_{ijk} \omega^{-1I}_j  \tilde{c}^I_k
	\right]  \label{eq:str-1} 
	\\
	\varepsilon^{(0)}_{\alpha} =& \frac{e^{\alpha}_i}{\tilde{l}}\left[u^{1J}_i - u^{1I}_i +  \frac{\partial u^{0J}_i}{\partial X_j} y^{I\!J}_j + \levicivita_{ijk}\left(\omega^{0J}_j + \varphi^{0J}_j+  \frac{\partial \omega^{-1J}_j}{\partial X_m} y^{I\!J}_m \right) \tilde{c}^J_k - \levicivita_{ijk}\left(\omega^{0I}_j+\varphi^{0I}_j\right) \tilde{c}^I_k\right] \label{eq:str0}
	\\
	\varepsilon^{(1)}_{\alpha} =&
	\frac{e^{\alpha}_i}{\tilde{l}}\left[ \frac{\partial u^{1J}_i}{\partial X_j} y^{I\!J}_j + \frac{1}{2}\frac{\partial^2 u^{0J}_i}{\partial X_j X_k} y^{I\!J}_j y^{I\!J}_k + \levicivita_{ijk}\left(\varphi^{1J}_j  +
	\frac{\partial \left(\omega^{0J}_j+\varphi^{0J}_j\right) }{\partial X_m} y^{I\!J}_m + \frac{1}{2}\frac{\partial^2 \omega^{-1J}_j}{\partial X_m X_n} y^{I\!J}_m y^{I\!J}_n \right) \tilde{c}^J_k - \levicivita_{ijk}\varphi^{1I}_j \tilde{c}^I_k\right] \label{eq:str1}
	\\
	\chi^{(-2)}_{\alpha} =&\frac{e^{\alpha}_i}{\tilde{l}}\left[\omega^{-1J}_i - \omega^{-1I}_i
	\right] \label{eq:curv-2}
	\\
	\chi^{(-1)}_{\alpha} =& \frac{e^{\alpha}_i}{\tilde{l}}\left[ \omega^{0J}_i+\varphi^{0J}_i + \frac{\partial \omega^{-1J}_i}{\partial X_j} y^{I\!J}_j - \omega^{0I}_i - \varphi^{0I}_i \right] \label{eq:curv-1}
	\\
	\chi^{(0)}_{\alpha} =& \frac{e^{\alpha}_i}{\tilde{l}}\left[\varphi^{1J}_i +
	\frac{\partial \left(\omega^{0J}_i+\varphi^{0J}_i\right) }{\partial X_j} y^{I\!J}_j + \frac{1}{2}\frac{\partial^2 \omega^{-1J}_i}{\partial X_j X_k} y^{I\!J}_j y^{I\!J}_k -\varphi^{1I}_i
	\right] \label{eq:curv0}
	\end{align}
\end{subequations}
The terms with higher $\eta$ power are omitted as they are negligibly small. 

\section{Taylor series of flux, traction, couple traction and moment of traction \label{ap:stress_taylor} }
The second nonzero components of stress-like variables are derived via Taylor series. The flux (or more precisely the permeability coefficient $\lambda$) is expanded around points $p=p^{(0)}_{\lambda}$, $\delta=\eta\delta^{(1)}_{\lambda}$, traction $\bt$ and moment of traction $\w$ around points $\bvarepsilon=\bvarepsilon^{(0)}$, $p=p^{(0)}_a$ and couple traction $\m$ around point $\bchi=\eta^{-1}\bchi^{(-1)}$, respectively.
\begin{subequations} \label{eq:stress1} 
	\begin{align} 
	\eta j^{(1)} &\approx - \eta p^{(1)}_{\lambda} g^{(0)} \frac{\partial{\lambda(p,\delta)}}{\partial p}\bigg|_
	{\begin{subarray}{l}p=p^{(0)}_{\lambda}\\\delta=\eta\delta^{(1)}_{\lambda}\end{subarray}}
	- \eta^2\delta^{(2)}_{\lambda} g^{(0)}\frac{\partial{\lambda(p,\delta)}}{\partial \delta}\bigg|_{\begin{subarray}{l}p=p^{(0)}_{\lambda}\\\delta=\eta\delta^{(1)}_{\lambda}\end{subarray}}  -\lambda\left(p^{(0)}_{\lambda}, \eta\delta^{(1)}_{\lambda} \right) \eta g^{(1)}
	\\
	&=\eta \frac{\partial j^{(0)}} {\partial p^{(0)}_{\lambda}} p^{(1)}_{\lambda}   
	+ \eta \frac{\partial j^{(0)}} {\partial \delta^{(1)}_{\lambda}} \delta^{(2)}_{\lambda} + \eta \frac{\partial j^{(0)}} {\partial g^{(0)}} g^{(1)} \nonumber
	\\
	\eta t^{(1)}_{\alpha} & \approx \eta \varepsilon^{(1)}_{\beta} \frac{\partial t_{\alpha}(\bvarepsilon,p)}{\partial \varepsilon_{\beta}}\bigg|_{\begin{subarray}{l} \bvarepsilon=\bvarepsilon^{(0)}\\p=p^{(0)}_{a}\end{subarray}}
	+
	\eta p^{(1)} \frac{\partial t_{\alpha}(\bvarepsilon, p)}{\partial p}\bigg|_{\begin{subarray}{l} \bvarepsilon=\bvarepsilon^{(0)}\\p=p^{(0)}_{a}\end{subarray}}
	= \eta \frac{\partial t^{(0)}_{\alpha}} {\partial \varepsilon^{(0)}_{\beta}}\varepsilon^{(1)}_{\beta} + \eta \frac{\partial t^{(0)}_{\alpha}} {\partial p^{(0)}_a}p^{(1)}_a 
	\\
	\eta^2 m^{(2)}_{\alpha} & \approx \chi^{(0)}_{\beta} \frac{\partial m_{\alpha}(\bchi)}{\partial \chi_{\beta}}\bigg|_{\bchi=\eta^{-1} \bchi^{(-1)}} = \eta^2 \frac{\partial m^{(1)}_{\alpha}} {\partial \chi^{(-1)}_{\beta}}\chi^{(0)}_{\beta} 
	\\
	\eta^2 w^{(2)}_i & \approx \eta \varepsilon^{(1)}_{\beta} \frac{\partial w_i(\bvarepsilon,p)}{\partial \varepsilon_{\beta}}\bigg|_{\begin{subarray}{l} \bvarepsilon=\bvarepsilon^{(0)}\\p=p^{(0)}_{a}\end{subarray}}
	+
	\eta p^{(1)} \frac{\partial w_i(\bvarepsilon, p)}{\partial p}\bigg|_{\begin{subarray}{l} \bvarepsilon=\bvarepsilon^{(0)}\\p=p^{(0)}_{a}\end{subarray}}
	= \eta^2 \frac{\partial w^{(1)}_i} {\partial \varepsilon^{(0)}_{\beta}}\varepsilon^{(1)}_{\beta} + \eta^2 \frac{\partial w^{(1)}_i} {\partial p^{(0)}_a}p^{(1)}_a 
	\end{align}
\end{subequations}

\section{Macroscale: from discrete to continuous representation \label{app:macro}}
Balance equations~\eqref{eq:balance0all} are summed over the whole RVE. The summations are straightforward in the case of fluxes (Eq.~\ref{eq:balance0fluxes}) and linear momentum (Eq.~\ref{eq:balance0forces}). In the case of angular momentum (Eq.~\ref{eq:balance0moments}), one must select some reference point in the global reference system $\X$ and sum all the moment contributions with respect to that particular point. It is convenient to choose the origin of reference system $\X$, therefore the distances to this reference point from an~arbitrary point $\x_r$ inside RVE will always have two components: $\X_r = \X_{\RVE} + \x_r$, the first component being the location of the RVE centroid in the global reference system, the latter is the position within the RVE. The moment due to inertia must be updated since the reference point is now changed to the origin of the $\X$ reference system. An~analogy to Eq.~\eqref{eq:inertia} with a~different reference point is used, $\uu^{(0)}$ is constant and $\sum_{I\in \VRVE} \rho V \left(\X_{\RVE} + \x_I + \rr_0\right)=\langle \rho \rangle \VRVE \X_{\RVE}$ because $\X_{\RVE}$ is the RVE centroid
\begin{align}
\sum_{I\in\VRVE} \M_{u\theta}^T \cdot \ddot{\uu}^{(0)} = 
\left[\sum_{I\in\VRVE} \rho V \llevicivita \cdot \left( \X_{\RVE} + \x_I + \rr_0\right)\right]\cdot \ddot{\uu}^{(0)} = \langle \rho \rangle \VRVE \llevicivita : \left( \X_{\RVE} \otimes \ddot{\uu}^{(0)} \right)
\end{align}
Also the moment due to external load $\bb$ acting in the RVE centroid is added. The three balance equations~\eqref{eq:balance0all} are transformed into
\begin{subequations} \label{eq:macroBalanceJFM}
	\begin{align}
	\frac{\eta}{\VRVE}\sum_{P\in \VRVE}\sum_{Q\in W} S^{\star} j^{(1)} &= \rho_{w0}\left[ \dot{v}_{c0}\left(1-b+\frac{p^{(0)}-p_0}{ K_{w}}\right)  +  v_{c0} \frac{ \dot{p}^{(0)}}{ K_w} + 3 b \dot{\varepsilon}^{(0)}_V +   \frac{\dot{p}^{(0)}}{ M_b }   \right] + q \label{eq:macroBalanceJFflow}
	\\
	\frac{\eta}{\VRVE}\sum_{I\in\VRVE}\sum_{J\in V}A^{\star} t^{(1)}_{\alpha}\e_{\alpha} &= \langle\rho\rangle \ddot{\uu}^{(0)} - \bb  \label{eq:macroBalanceJFforces}
	\\
	\frac{\eta^2}{\VRVE}\sum_{I\in\VRVE}\sum_{J\in V} A^{\star}\left(\w^{(2)} + m^{(2)}_{\alpha} \e_{\alpha} \right) &=  \langle \rho \rangle \llevicivita : \left( \X_{\RVE} \otimes \ddot{\uu}^{(0)} \right) - \llevicivita:\left(\X_{\RVE} \otimes \bb\right) \label{eq:macroBalanceMomentsA}
	\end{align}
\end{subequations}
New variable $v_{c0}$ denoting the crack volume density in the whole RVE is introduced into the first equation~\eqref{eq:macroBalanceJFflow} for sake of simplicity 
\begin{align} \label{eq:RVEcracks}
v_{c0} = \frac{1}{\VRVE}\sum_{P\in \VRVE}\sum_{Q\in W} v^{(0)}_c W = \langle v^{(0)}_c \rangle
\end{align}

Equations~\eqref{eq:stress1} are now substituted into the balance equations~\eqref{eq:macroBalanceJFM}
\begin{subequations} \label{eq:macroBalance1}
	\begin{align}
	\frac{\eta}{\VRVE}\sum_{P\in \VRVE}\sum_{Q\in W} S^{\star} \left( \frac{\partial j^{(0)}} {\partial p^{(0)}} p^{(1)}_{\lambda}
	+ \frac{\partial j^{(0)}} {\partial \delta^{(0)}} \delta^{(1)}_{\lambda} + \frac{\partial j^{(0)}} {\partial g^{(0)}} g^{(1)} \right) =&\rho_{w0}\left[ \dot{v}_{c0}\left(1+b+\frac{p^{(0)}-p_0}{ K_{w}}\right) \right.
	\\ & \left.  +  v_{c0} \frac{ \dot{p}^{(0)}}{K_w} + 3 b \dot{\varepsilon}^{(0)}_V +   \frac{\dot{p}^{(0)}}{ M_b }   \right] + q
	\\
	\frac{\eta}{\VRVE}\sum_{I\in\VRVE}\sum_{J\in V}A^{\star} \left( 
	\frac{\partial t^{(0)}_{\alpha}} {\partial \varepsilon^{(0)}_{\beta}} \varepsilon^{(1)}_{\beta} + \frac{\partial t^{(0)}_{\alpha}} {\partial p^{(0)}} p^{(1)}_a\right)\e_{\alpha} =& \langle\rho\rangle \ddot{\uu}^{(0)} - \bb 
	\\
	\frac{\eta^2}{\VRVE}\sum_{I\in\VRVE}\sum_{J\in V} A^{\star}\left(
	\frac{\partial \w^{(1)}} {\partial \varepsilon^{(0)}_{\beta}} \varepsilon^{(1)}_{\beta}
	-\frac{\partial \w^{(1)}} {\partial p^{(0)}} p^{(1)}_a + 
	\frac{\partial m^{(1)}_{\alpha}} {\partial \chi^{(-1)}_{\beta}} \chi^{(0)}_{\beta} \e_{\alpha} \right) =& \llevicivita:\left[\X_{\RVE} \otimes \left( \left\langle\rho\right\rangle \ddot{\uu}^{(0)}-\bb\right)\right]
	\end{align}
\end{subequations}
Each element on the left-hand side appears in the summation twice, once connecting node $I$ with node $J$ and once connecting $J$ with $I$ (or $P$ with $Q$, respectively).  The contact vectors of these two identical element have opposite directions ($\x_{I\! J}=-\x_{J\! I}$, $\x_{PQ}=-\x_{QP}$). Also their directional vectors have opposite sign ($^{I\!J\!\!}\e_{\alpha} = -^{J\!I\!}\e_{\alpha}$). The tractions $t^{(0)}_{\alpha}$, strains $\varepsilon^{(0)}_{\alpha}$, pressures $p^{(-1)}$, average pressures $p^{(1)}_a$ and $p^{(1)}_{\lambda}$, normal crack openings $\delta^{(0)}_{\lambda}$ and $\delta^{(1)}_{\lambda}$, couple tractions $m^{(1)}_{\alpha}$ and curvatures $\chi^{(-1)}_{\beta}$ are identical, moments of traction $w^{(1)}$ as well as fluxes $j^{(-1)}$ and pressure gradients $g^{(-1)}$ have the same magnitude but opposite direction. 

Three terms with $p^{(1)}$ and one with $\delta^{(1)}_{\lambda}$ disappear from Eqs.~\eqref{eq:macroBalance1} as the same expressions are subtracted. The remaining terms lead to summation of two pressure gradients and subtraction of strains and curvatures of those two identical elements, which according to Eqs.~\eqref{eq:g1}, \eqref{eq:str1} and \eqref{eq:curv0} read (using $\bomega^{(-1)}=\bm{0}$; $p^{(0)}$, $\uu^{(0)}$ and $\bvarphi^{(0)}$ are constants)
\begin{subequations} \label{eq:summation}
	\begin{align}
	^{PQ\!}g^{(1)} + ^{QP\!\!\!\!}g^{(1)} &= \frac{1}{\tilde{h}} \left[\frac{\partial \left(p^{1Q} - p^{1P}\right)}{\partial X_i}  y^{PQ}_i + \frac{\partial^2 p^{(0)}}{\partial X_i X_j} y^{PQ}_i y^{PQ}_j \right] = \frac{\partial g^{(0)}}{\partial X_i} y^{PQ}_i
	\\
	^{I\!J\!}\varepsilon^{(1)}_{\beta} - ^{J\!I\!\!}\varepsilon^{(1)}_{\beta} =& \frac{e^{\beta}_i}{\tilde{l}}\left[ \frac{\partial \left(u^{1J}_i-u^{1I}_i\right)}{\partial X_j} y^{I\!J}_j + \levicivita_{ijk} \left( \frac{\partial \left(\omega^{0J}_j+\varphi^{(0)}_j\right) }{\partial X_m} \tilde{c}^J_k  - \frac{\partial \left(\omega^{0I}_j+\varphi^{(0)}_j\right)}{\partial X_m}  \tilde{c}^I_k\right)y^{I\!J}_m \right. \nonumber
	\\ & \left. + \frac{\partial^2 u^{(0)}_i}{\partial X_j X_k} y^{I\!J}_j y^{I\!J}_k  \right] = \frac{\partial \varepsilon^{(0)}_{\beta} }{\partial X_i}y^{I\!J}_i \label{eq:summationb}
	\\ 
	^{I\!J\!}\chi^{(0)}_{\beta} - ^{J\!I\!\!}\chi^{(0)}_{\beta} =& \frac{e^{\beta}_i}{\tilde{l}}\left[ \frac{\partial \left(\omega^{0J}_i+\varphi^{(0)}_i - \omega^{0I}_i - \varphi^{(0)}_i \right) }{\partial X_j} y^{I\!J}_j \right]  = \frac{\partial \chi^{(-1)}_{\beta} }{\partial X_i}y^{I\!J}_i  
	\end{align} 
\end{subequations}

Equations~\eqref{eq:summation} are substituted back to the Eqs.~\eqref{eq:macroBalance1} along with application of the chain rule, replacement of the double summation to single summation over all the elements and change from $\y_{I\!J}$ to $\x_{I\!J}/\eta$ to obtain
\begin{subequations}\label{eq:macroBalance2}
	\begin{align}
	\frac{1}{\VRVE}\sum_{d\in \VRVE} S^{\star} \frac{\partial j^{(0)}} {\partial X_i} x^{PQ}_i & = \rho_{w0}\left[ \dot{v}_{c0}\left(1-b+\frac{p^{(0)}-p_0}{ K_{w}}\right)  +  v_{c0} \frac{ \dot{p}^{(0)}}{K_w} + 3 b \dot{\varepsilon}^{(0)}_V +   \frac{\dot{p}^{(0)}}{ M_b }   \right] + q  \label{eq:macroBalance2flux}
	\\
	\frac{1}{\VRVE}\sum_{e\in\VRVE} A^{\star} 
	\frac{\partial t^{(0)}_{\alpha}} {\partial X_i} x^{I\!J}_i
	\e_{\alpha} & = \langle\rho\rangle \ddot{\uu}^{0I} - \bb  \label{eq:macroBalance2force}
	\\
	\frac{\eta}{\VRVE}\sum_{e\in\VRVE} A^{\star}\left(
	\frac{\partial \w^{(1)}} {\partial X_i} x^{I\!J}_i + 
	\frac{\partial m^{(1)}_{\alpha}} {\partial X_i} x^{I\!J}_i \e_{\alpha} \right) & =  \llevicivita:\left(\X_{\RVE} \otimes \left( \left\langle\rho\right\rangle \ddot{\uu}^{(0)}-\bb\right)\right) \label{eq:macroBalance2moment}
	\end{align}
\end{subequations}

The expression for derivative of the moment of traction $\eta\w^{(1)}= t^{(0)}_{\alpha}\llevicivita:(\X_c \otimes \e_{\alpha})$ is modified by substituting the position of the integration point $\X_c = \X_{\RVE}+\x_c$
\begin{align}
\eta\frac{\partial \w^{(1)} } {\partial X_n} = 
\levicivita_{ijk}\frac{\partial \left( t^{(0)}_{\alpha}  (X^{\RVE}_j+x^c_j) e^{\alpha}_k  \right)} {\partial X_n} &=
\levicivita_{ijk} \frac{\partial \left( t^{(0)}_{\alpha}  x^c_j e^{\alpha}_k  \right)} {\partial X_n} + \levicivita_{ijk} \frac{\partial  t^{(0)}_{\alpha}}{\partial X_n} X^{\RVE}_j  e^{\alpha}_k + \levicivita_{ink}  t^{(0)}_{\alpha} e^{\alpha}_k 
\end{align}
Thanks to periodicity of the RVE geometry providing $\partial x^{I\!J}_i/\partial X_j = \partial e^{\alpha}_i/\partial X_j = \partial A^{\star}/\partial X_j= 0$, variables $x_{I\!J}$, $\e_{\alpha}$ and $A^{\star}$ can be moved inside or outside the differentiation with respect to $\X$. 
%\begin{align}
%A^{\star}\frac{\partial \bullet } {\partial X_n}x^{I\!J}_n \e_{\alpha} = \frac{\partial \left( A^{\star} \bullet x^{I\!J}_n \e_{\alpha} \right)} {\partial X_n}
%\end{align}
It allows  one to define the following quantities called total macroscopic stress and total couple stress tensors 
\begin{subequations} \label{eq:totalmacrostress}
	\begin{align}
	\bsigma_{\mathrm{t}} &= \frac{1}{\VRVE}\sum_{e\in\VRVE}  lA^{\star} t^{(0)}_{\alpha} \e_{N} \otimes \e_{\alpha}
	\\
	\bmu_{\mathrm{t}} &= \frac{1}{\VRVE}\sum_{e\in\VRVE}  lA^{\star} \e_{N} \otimes \left( \eta m^{(1)}_{\alpha}  \e_{\alpha} + t^{(0)}_{\alpha} \llevicivita : \left( \x_c \otimes \e_{\alpha}\right) \right)
	\end{align} 
\end{subequations}
and rewrite the mechanical balance equations \eqref{eq:macroBalance2force} and \eqref{eq:macroBalance2moment} accordingly
\begin{subequations} \label{eq:mechanicsprefinal}
	\begin{align}
	\nabla_X\cdot \bsigma_{\mathrm{t}} &=  \langle\rho\rangle \ddot{\uu}^{(0)} - \bb
	\\
	\nabla_X\cdot \bmu_{\mathrm{t}} + \llevicivita:\bsigma_{\mathrm{t}} &= \llevicivita:\left[\X_{\RVE} \otimes \left( \left\langle\rho\right\rangle \ddot{\uu}^{(0)}-\bb - \nabla_X\cdot \bsigma_{\mathrm{t}}\right)\right]
	\end{align}
\end{subequations}
The last term of the second equation is zero because its inner part exactly correspond to the first equation. Equations~\eqref{eq:macroBalance2flux}, \eqref{eq:totalmacrostress} and \eqref{eq:mechanicsprefinal} can now be easily rewritten to Eqs.~\eqref{eq:macrofinal} and \eqref{eq:finalmacrostress}.

\section{Volumetric strain expansion \label{ap:volumetric_strain_expansion}}
Equation~\eqref{eq:balance0fluxes} features term $\varepsilon_V^{(0)}$. Volumetric strain is defined in Sec.~\eqref{model} as one third of relative change of the tetrahedron volume. Let us denote four vertices of tetrahedron as $a$--$d$, areas and outward normals of faces opposite to these vertices as $A_a$--$A_d$ and $\n_a$--$\n_d$, respectively. The volumetric strain can be calculated as 
\begin{align}
\varepsilon_V &= - \frac{1}{9V}\sum_{I=a\dots d} A_I \uu_I \cdot \n_I
\end{align}
The negative sign is included since the normals are in the outward  direction. Re-introducing the expansions from Eqs.~\eqref{eq:expansion_u}, \eqref{eq:Taylorseries_u} and \eqref{eq:vol_strain_exp}, one arrives at
\begin{align}
\varepsilon_V^{(0)} &= - \frac{1}{9\tilde{V}}\sum_{I=a\dots d} \tilde{A}_I  u^{1I}_i n^I_i - \hat{\varepsilon}_V^{(0)} &  \mbox{with} && \hat{\varepsilon}_V^{(0)} & = \frac{1}{9\tilde{V}}\sum_{I=a\dots d} \tilde{A}_I \frac{\partial u^{(0)}_i}{\partial X_j} y^{dI}_j n^I_i
\end{align} 
where we randomly chose one of the vertices as the primary one ($d$) and use it in the Taylor series to estimate displacements in other vertices. 

Let us now further focus on the eigen part coming from the macroscopic displacement gradient. Since the gradient of displacements is constant over the RVE, it can be moved in front of the summation. The normal $\n_a$ can be computed as the normalized cross product of vectors pointing from $d$ to $b$ and $c$: $n^a_i = \levicivita_{ilm} y^{db}_l y^{dc}_m/2\tilde{A}_a$. One therefore reads
\begin{align}
\hat{\varepsilon}_V^{(0)} = \frac{1}{9\tilde{V}} \frac{\partial u^{(0)}_i}{\partial X_j} \sum_{I\in\{a,b,c\}} y^{dI}_j n^I_i A_I = \frac{1}{18\tilde{V}}\frac{\partial u^{(0)}_i}{\partial X_j} \delta_{ij}  \levicivita_{klm} y^{da}_k y^{db}_l y^{dc}_m = -\frac{1}{3} \frac{\partial \uu^{(0)}}{\partial \X}:\bm{1}  = \frac{1}{3}\mathrm{tr}\left(\nabla \uu^{(0)}\right)
\end{align}
The second equality substitutes the normal $\n_I$ and re-arranges the multiplications and the third equality uses well known formula for the tetrahedron volume based on the determinant $\levicivita_{klm} y^{da}_k y^{db}_l y^{dc}_m = -6\tilde{V}$. The eigen part of volumetric strain becomes  one third of the macroscopic displacement gradient trace, i.e.,~the volumetric part of the macroscopic strain.
\section{Constitutive relations \label{app:A}}

The mechanical constitutive model relates mechanical strain vector $\bvarepsilon$ to  traction in the solid, $\s=f_s(\bvarepsilon)$. It is formulated in local coordinate system distinguishing normal ($N$) and two tangential ($M$ and $L$) directions. The formulation is adopted from Ref.~\parencite{CusCed07} and simplified by omitting the confinement effect, assuming non-decreasing damage also in compression and reducing number of material parameters to 4: normal elastic constant $E_0$, tangential/normal stiffness ratio $\alpha$, tensile strength $\ft$ and tensile fracture energy $\Gt$. 

Normal and tangential tractions read
\begin{align}  
s_N &=  (1-d) E_0 \varepsilon_N & s_M &=  (1-d) E_0 \alpha \varepsilon_M & s_L &=  (1-d) E_0 \varepsilon_L 
\end{align} 
$d$ is non-decreasing damage parameter ranging from 0 (healthy material) to 1 (completely damaged material). If increasing, it is updated  via equivalent boundary traction, $\seff$, and equivalent strain, $\eeff$, as $d = 1-\frac{\seff}{E_0\eeff}$. Equivalent strain is defined as $\eeff=\sqrt{\varepsilon_N^2+\alpha\varepsilon_T^2}$, $\varepsilon_T= \sqrt{\varepsilon_M^2+\varepsilon_L^2}$. Let us also define direction of straining $\omega$ evaluated from $\tan \omega=\varepsilon_N / \sqrt{\alpha} \varepsilon_T$.
The equivalent boundary traction reads
\begin{align}  
\seff =  \feff \exp\left( \frac{K}{\feff}\left\langle \chi -\frac{\feff}{E_0}\right\rangle\right) \label{eq:seq}
\end{align} 
The angled brackets return the positive part, $\feff$ denotes equivalent strength, $K$ is the initial slope in the nonlinear regime and $\chi$ represents the loading history
\begin{align}
\feff &= \begin{cases} \frac{16\ft}{\sqrt{\sin^2\omega+\alpha\cos^2\omega}} & \omega<\omega_0 \\ \ft \frac{4.52 \sin\omega-\sqrt{20.0704 \sin^2\omega+9 \alpha\cos^2\omega }}{0.04\sin^2\omega-\alpha \cos^2\omega} & \omega\geq\omega_0 \end{cases} \label{eq:feq}
\\
K &= \begin{cases} 0.26E_0\left[ 1-\left( \frac{\omega+\pi/2}{\omega_0+\pi/2}\right)^2\right] & \omega<\omega_0 \\ -K_{\mathrm{t}}\left[ 1-\left( \frac{\omega-\pi/2}{\omega_0-\pi/2}\right)^{n_t}\right] & \omega\geq\omega_0 
\end{cases}
\\
\chi &= \begin{cases} \eeff & \omega<\omega_0 \\ \eeff\frac{\omega}{\omega_0} + \sqrt{\max\varepsilon_N^2 + \alpha\max\varepsilon_T^2}\left(1-\frac{\omega}{\omega_0}\right) & \omega_0\leq\omega<0 \\ \sqrt{\max\varepsilon_N^2 + \alpha\max\varepsilon_T^2} & \omega\geq 0
\end{cases}
\end{align} 
with $\omega_0$ being the direction at which both branches of Eq.~\eqref{eq:feq} are equal.

\begin{table}
	\begin{center}
		\begin{tabular}{crcrl}	
			&material parameter & symbol & value & units \\\hline
			\multirow{4}{*}{mechanics}&normal elastic constant & $E_0$& 21.5 & GPa \\ 
			&tangential/normal stiffness ratio & $\alpha$ & 0.3 & -\\
			&tensile strength & $\ft$ & 2.1 & MPa\\
			&tensile fracture energy & $\Gt$ & 50 &  J/m$^2$\\\hline
			\multirow{5}{*}{\pbox{20cm}{mass\\transport}}&capacity & $c$& 1.62$\times$10$^{-8}$ & s$^2$/m$^2$ \\ 
			&density & $\rho_{w0}$ & 1000 & kg/m$^3$\\
			&permeability & $\kappa$ & 5$\times$10$^{-18}$ & m$^2$\\
			&viscosity & $\mu$ & 8.9$\times$10$^{-4}$ &  Pa$\cdot$s\\
			&reference pressure & $p_0$ & 0 or 1 & MPa\\\hline
			\multirow{3}{*}{coupling}&crack tortuosity & $\xi$ & 1 & -\\
			&bulk modulus & $K_w$ & 2.15 & GPa\\
			&Biot coefficient & $b$ & 0--1 & -\\\hline
		\end{tabular}
	\end{center}
	\caption{Material parameters used in the verification examples. Approximate values of macroscopic elastic modulus, $E$, and Poisson's ratio, $\nu$, are estimated by analyses of the RVE response as $13.97$\,GPa and $0.175$, respectively. Macroscopic RVE permeability coefficient of the intact material is exactly equal to the microscopic one, i.e.,~$\rho_{w0}\kappa/\mu = 5.618\times10^{-12}$\,s.}
	\label{tab:materparams}
\end{table}

The initial slope of the strain softening, $K$, is defined using $K_{\mathrm{t}}$ and $K_{\mathrm{s}}$, the slopes for pure tension and shear, respectively. These are dependent on contact length $l$ and read
\begin{align}
K_{\mathrm{t}} &= \frac{2E_0\ft^2 l}{2E_0\Gt - \ft^2l} & K_{\mathrm{s}} &= \frac{18\alpha E_0\ft^2 l}{32\alpha E_0\Gt - 9\ft^2l}
\end{align}
Finally, parameter $n_t$ reads
\begin{equation}
n_t = \frac{\ln\left(K_{\mathrm{t}}/(K_{\mathrm{t}}-K_{\mathrm{s}})\right)}{\ln\left( 1-2\omega_0/\pi\right)}
\end{equation}

If the confinement effect is included, the constitutive model features also the volumetric strain, which, after expansion, would appear through its macroscopic (slow) component and microscopic (fast) component evaluated at the RVE from Delaunay tetrahedrons according to Appendix~\ref{ap:volumetric_strain_expansion}.

Transport constitutive equation relates the scalar flux, $j$, and the scalar pressure gradient, $g$: $j=f_{j} (p,g,\delta_N)$. It is assumed to be in a~form $j=-\lambda(p, \delta_N) g$, flux is linearly dependent on pressure gradient. $\lambda$ is material permeability coefficient dependent on pressure $p$ and crack opening $\delta_N$. For sake of simplicity, the effect of pressure is not considered here even though the homogenization derivation assumes such dependency. Reader is referred to paper~\parencite{EliYin-22} where the pressure dependent permeability coefficient is used in the studies with the \emph{homogenized} and \emph{full} model. The crack opening dependence is adopted from Ref.~\parencite{GraBol16} as a~summation of permeability coefficient of intact material and an~effect of cracks 
\begin{align}
\lambda = \frac{\rho_{w0}\kappa}{\mu} + \frac{\xi\rho_{w0}}{12\mu S}\sum_{i=1}^3 \delta_{Ni}^3 l_{\mathrm{c}i}
\end{align}
$\rho_{w0}$ is fluid density, $\kappa$ is permeability, $\mu$ is viscosity, $\xi$ is crack tortuosity parameter, $\delta_N$ and $l_{\mathrm{c}}$ are normal crack openings and crack lengths of associated mechanical elements, see Ref.~\parencite{GraBol16} for details.

Material parameters used throughout the paper ale listed in Tab.~\ref{tab:materparams}. These parameters were partly taken from literature and partly identified by simulating Brazilian test followed by measurement of water flux  through the cracked specimens according to Refs.~\parencite{WanJan-97,AldSha-99}.

\section{Analytical solution to the Terzaghi's consolidation problem \label{app:B}}
Analytical solutions~\parencite{Ter23,DetChe93} to the elastic Terzaghi's consolidation problems (pressure and traction loading) are the following
\begin{align}
\mathrm{loading\ by\ pressure} & & p(\chi, \tau) &= p^{\star} F_1(\chi, \tau) & u(\chi, \tau) &= -\frac{p^{\star}\Upsilon L}{G} F_2(\chi, \tau)
\\
\mathrm{loading\ by\ traction} & & p(\chi, \tau) &= \frac{t_x^{\star}\Upsilon L}{GS} \left[1-F_1(\chi, \tau)\right] & u(\chi, \tau) &= t_x^{\star}L\left[\frac{(1-2\nu_u)(1-\chi)}{2G(1-\nu_u)}
+ \frac{(\nu_u-\nu)F_2(\chi, \tau)}{2G(1-\nu)(1-\nu_u)}\right]
\end{align}
where functions $F_1$ and $F_2$ read
\begin{align}
F_1(\chi, \tau) &= 1-\sum_{m=1,3,\dots}^{\infty} \frac{4}{m\pi}\sin\left(\frac{m\pi\chi}{2}\right)\exp\left(-m^2\pi^2\tau\right) \\
F_2(\chi, \tau) &= \sum_{m=1,3,\dots}^{\infty} \frac{8}{m^2\pi^2}\cos\left(\frac{m\pi\chi}{2}\right)\left[1-\exp\left(-m^2\pi^2\tau\right)\right]
\end{align}
$\chi=x/L$, $\tau=\lambda t/4CL^2$, $\Upsilon=b(1-2\nu)/[2(1-\nu)]$, $L=\mathrm{prism\ length}=0.5$\,m, $G=E/2(1+\nu)$, $C=(1-\nu_u)(1-2\nu)/[M_b (1-\nu)(1-2\nu_u)]$, $M_b=1/c$, $\nu_u=(3K_u-2G)/2(3K_u+G)$ and $K_u=M_b b^2 + E/3(1-2\nu)$. 

\section{Analytical solution to the elastic pressurized hollow cylinder problem \label{app:C}}
\textcite{GraFah-15} derived analytical solution for the steady state elastic problem in terms of pressure $p$ and radial displacement $u_{\mathrm{r}}$. 
\begin{align}
p ( r )&= p_{\mathrm{i}}\frac{\ln \frac{r_{\mathrm{o}}}{r} }{\ln \frac{r_{\mathrm{o}}}{r_{\mathrm{i}}}}\\
u_{\mathrm{r}}(r) &= -\frac{p_{\mathrm{i}}b}{E} \frac{1-\nu^2}{2}\left[\frac{r_{\mathrm{o}}^2}{r_{\mathrm{o}}^2-r_{\mathrm{i}}^2}\left(\frac{r_{\mathrm{o}}^2(1+\nu)}{r(1-\nu)} + r  \right) + r\frac{\frac{1}{1+\nu} - \ln\frac{r}{r_{\mathrm{i}}}}{\ln \frac{r_{\mathrm{o}}}{r_{\mathrm{i}}}}  \right] - \frac{p_{\mathrm{i}}(1-b)}{E}\frac{r_{\mathrm{o}}^2r_{\mathrm{i}}^2}{r_{\mathrm{o}}^2-r_{\mathrm{i}}^2}\left(\frac{1+\nu}{r} + \frac{r (1-\nu)}{r_{\mathrm{o}}^2}\right)
\end{align}
$r$ is the radial coordinate, $p_i$ is the internal pressure, $r_{\mathrm{i}}$ and $r_{\mathrm{o}}$ are internal and external radii.

\section*{Acknowledgements}
Jan Eliáš gratefully acknowledges financial support from the Czech Science Foundation under project no. GA19-12197S.

\printbibliography[heading=bibintoc]

\end{document}